\documentclass[12pt]{article}
\usepackage[dvipdfmx,hiresbb]{graphicx}
\usepackage{float}
\begin{document}
\global\arraycolsep=2pt 
\newcommand{\p}{{\bf p}}
\newcommand{\q}{{\bf q}}
\newcommand{\s}{{\bf s}}
\newcommand{\x}{{\bf x}}
\newcommand{\y}{{\bf y}}
\thispagestyle{empty} 
\begin{titlepage}    
\begin{flushright}
TNCT-1401
\end{flushright}
\vspace{1cm}
\topskip 3cm
\begin{center}
  \Large{\bf 
     Analytically expressed constraint on two Majorana phases
     in neutrinoless double beta decay
     }  \\
\end{center}                                   
\vspace{0.6cm}
\begin{center}
Shinji Maedan            
  \footnote{ E-mail: maedan@tokyo-ct.ac.jp}   
           \\
\vspace{0.8cm}
  \textsl{ Department of Physics, Tokyo National College of Technology,
        Kunugida-machi,Hachioji-shi, Tokyo 193-0997, Japan}
\end{center}                                               
\vspace{0.5cm}
\begin{abstract}
\noindent       
\\
We assume that neutrinoless double beta decay is caused
   by the exchange of three light Majorana neutrinos.
Under this assumption, we obtain, by the method of perturbation,
   the equation representing the isocontour
   of effective Majorana mass
   which is the function of two CP-violating Majorana phases.
The equation representing the isocontour (constraint equation between
   two Majorana phases) is expressed analytically by six parameters:
   two lepton mixing angles, two kinds of neutrino mass squared differences,
   lightest neutrino mass scale, and the effective Majorana mass.
We discuss how the constraint equation between two Majorana phases changes
   when the lightest neutrino mass scale is varied.
\end{abstract} 
\end{titlepage}
%
%
\setcounter{page}{1}
\section{Introduction}
At present, it is unknown whether the neutrinos are massive Dirac particles
   or massive Majorana particles.
If neutrinoless double beta decay ($0 \, \nu \beta \beta$) is  caused
   by the exchange of three light Majorana neutrinos, the observation of
   $0 \, \nu \beta \beta$ is the evidence that the neutrinos are Majorana fermions \cite
   {rf:DoiKotTak,rf:BilPet,rf:Rod}.
In the process of neutrinoless double beta decay, 
\begin{equation}
   (A, Z) \rightarrow  (A, Z+2) + e^{-} + e^{-},    
  \label{ac}
\end{equation}
the lepton number is not conserved.
Neutrinoless double beta decay has not been observed yet.
Many experiments of  $ 0 \, \nu \beta \beta$ are in progress and planed:
    CANDLES  \cite{rf:UmeKisOga}, NEMO-3 \cite{rf:Arg}, SOLOTVINO \cite{rf:Dan},  
    CUORICINO \cite{rf:Arn}, EXO-200 \cite{rf:Aug}, 
    KamLAND-Zen \cite{rf:Gan}, etc.
If neutrinos are Majorana fermions, lepton mixing matrix (MNS matrix \cite{rf:MakNakSak})
   $U$ is represented by six parameters: three lepton mixing angles
   $(\theta_{12}, \theta_{23},  \theta_{13})$, CP-violating Dirac phase $\delta$,
   and two CP-violating Majorana phases $\alpha, \beta$.
These $\alpha$ and $ \beta$ are the degrees of freedom of phase which come from
   the assumption that neutrinos are Majorana fermions \cite{rf:BilHosPet}.
Among these six parameters in the lepton mixing matrix, three parameters
    $\theta_{12}$, $ \theta_{23}$, and $  \theta_{13} $ are
     measured by experiments \cite{rf:Gon}.
The measurement of $  \theta_{13} $ especially was done recently \cite{rf:Abe},
   whose effect on the study of  $ 0 \, \nu \beta \beta$ had been
   investigated \cite{rf:LinMerRod}.
Dirac phase $\delta$ and two Majorana phases $\alpha$ and $\beta$ are unknown
   by experiments.
If one assumes that neutrinoless double beta decay is caused by the exchange of
   three light Majorana neutrinos, the amplitude of $ 0 \, \nu \beta \beta$ is proportional
   to the effective Majorana mass $ \vert m_{ee} \vert $,
\begin{equation}
    \vert m_{ee} \vert
   =  \left\vert \, m_1  \vert U_{e 1} \vert^2 + m_2 \vert U_{e 2} \vert^2  e^{2 i \alpha}
    +  m_3 \vert U_{e 3} \vert^2  e^{2 i \beta} \,  \right\vert,
  \label{aa}
\end{equation}
where $m_1, m_2$, and $m_3$ are mass of three neutrinos, and $U$ is the lepton
   mixing matrix.
There are two possibilities of neutrino mass spectrum: the normal mass
   ordering $(m_3 >m_2>m_1)$ and the inverted mass ordering $(m_2 >m_1>m_3)$.

The effective Majorana mass $ \vert m_{ee} \vert $ depends on the seven parameters:
   mixing angles $(\theta_{12}, \theta_{13})$, neutrino mass $(m_1, m_2, m_3)$, 
   and Majorana phases $(\beta, \alpha)$.
As regarding these seven parameters,
   mass squared differences
   $ \bigtriangleup m_{\odot}^2   \equiv  m_2^2 - m_1^2 $ and
   $ \bigtriangleup  m_{\rm A}^2  \equiv  \vert m_3^2 - m_1^2 \vert
                           \approx \vert m_3^2 - m_2^2 \vert $,
   $\theta_{12}$, and $ \theta_{13}$
   are measured by experiments \cite{rf:Gon},
   whereas the absolute neutrino mass scale and Majorana phases $\beta, \alpha$
   (if Majorana particles) are not measured.
While information on Majorana phases $\beta, \alpha$ is obtained by neutrinoless
   double beta decay experiments \cite{rf:MatTakFuk,rf:Rod2,rf:MatTakFukNis,
   rf:Rod3,rf:MinNunQui},
   it is notable that the other experiments measuring
   the absolute neutrino mass scale are also important to determine the phases
   $\beta, \alpha$ as discussed in  Ref.\cite{rf:MinNunQui}.
In studying neutrinoless double beta decay, people often use the method in which
   one regards the lightest neutrino mass as a free parameter and analyzes the
   effective Majorana mass $\vert m_{ee} \vert$ for each given value of the lightest
   neutrino mass.
The effective Majorana mass $\vert m_{ee} \vert$ is considered to be a function of two
   variables $\beta$ and $ \alpha$ provided that the value of the lightest neutrino mass 
   is given by hand.
Even if the value of $\vert m_{ee} \vert$ is obtained by $ 0 \, \nu \beta \beta$ experiments,
   we can not determine both values of $\beta$ and $ \alpha$ simultaneously
   for a given value of the lightest neutrino mass.
Instead, we can obtain the constraint between the Majorana phases,
   $\beta$ and $ \alpha$ .
Before $\theta_{13}$ was measured by the experiments \cite{rf:Abe},
   the constraint between $\beta$ and $ \alpha$ had been examined for given values
   of $\vert m_{ee} \vert$ and neutrino mass $  m_i \, (i=1,2,3) $ by numerical
   calculations \cite{rf:MinYas,rf:Rod2,rf:MatTakFukNis,rf:NunTevFun,rf:AbaBha}.
In the numerical calculations, the authors gave some values to the parameters,
   $\theta_{12}, \theta_{13}$, $ \bigtriangleup m_{\odot}^2,  \bigtriangleup  m_{\rm A}^2 $,
   and the lightest neutrino mass, respectively.
   
In this paper, we obtain the analytic equation representing the isocontour of the effective
   Majorana mass $\vert m_{ee} \vert$ on the $ \beta \alpha-$plane for each of the
   given values of the lightest neutrino mass.
The equation of the isocontour of  $\vert m_{ee} \vert$ (constraint between
   $\beta$ and $ \alpha$) is derived by the method of perturbation for the normal mass
   ordering case and the inverted mass ordering case, respectively.
The equation is represented analytically by the six parameters:
   $\theta_{12}, \theta_{13}$, $ \bigtriangleup m_{\odot}^2,  \bigtriangleup  m_{\rm A}^2 $,
   the lightest neutrino mass scale, and of course $\vert m_{ee} \vert$.
Because the effective Majorana mass is invariant under
   $ \beta \rightarrow \beta + n \pi,
       ~  \alpha \rightarrow \alpha + m \pi, (n,m \in {\bf Z})$,
\begin{equation}
   \vert m_{ee} \vert (\beta, \alpha) 
   =   \vert m_{ee} \vert (\beta + n \pi, \alpha + m \pi),
  \label{ab}
\end{equation}
the isocontour of $ \vert m_{ee} \vert $ around a point $(\beta, \alpha)$ in the
   $\beta \alpha-$plane is the same as that
   around the point $(\beta + n \pi, \alpha + m \pi) $.
If neutrinoless double beta decay is observed, the next challenging task is to restrict
   the values of the Majorana phases $ \beta$ and/or $\alpha$.
At that stage, the analytically expressed constraint between $ \beta$ and $\alpha$
   is useful.

The paper is organized as follows.
In section 2, the notations we use are introduced, and the characteristic features of
   the normal mass ordering and the inverted mass ordering are described, respectively.
In section 3, the isocontour of the effective Majorana mass $ \vert m_{ee} \vert $
   in the $ \beta \alpha-$plane (constraint between $ \beta$ and $\alpha$) is obtained
   in the case of the normal mass ordering.
To describe more concretely, the isocontour of $ \vert m_{ee} \vert $ around the point
   of maximum $ \vert m_{ee} \vert $ in  the $ \beta \alpha-$plane, that around the point
   of minimum $ \vert m_{ee} \vert $, and that except around the point of maximum or
   minimum $ \vert m_{ee} \vert $ are obtained by the method of perturbation.
In section 4, the same kind of things as section 3 is described in the case of 
   the inverted mass ordering.
Section 5 is devoted to conclusions.
In Appendix, the perturbative method used in section 3 is explained.
%
%
%
%
\section{Formalism}
If we assume that neutrinoless double beta decay ($ 0 \, \nu \beta \beta$) is generated
   by the exchange of
   the three Majorana neutrinos with light mass, the amplitude of $ 0 \, \nu \beta \beta$
   is proportional to the effective Majorana mass $  \vert m_{ee} \vert $,
\begin{equation}
    \vert m_{ee} \vert
   = \vert  \sum_{i=1}^{3} m_i \, U_{e i}^2  \vert
   =  \left\vert \, m_1  U_{e 1}^2 + m_2  U_{e 2}^2
    +  m_3  U_{e 3}^2  \right\vert,
  \label{ba}
\end{equation}
where $m_i (i=1,2,3)$ is neutrino mass of $i-$th mass eigenstate.
The unitary matrix $U$ is the lepton mixing matrix (MNS matrix) and parametrized
   as follows \cite{rf:LinMerRod},
\begin{eqnarray}
  U =  \left(   \begin{array}{ccc}
                 1  &        0       &   0           \\
                 0  &   c_{23}   &   s_{23}    \\
                 0  &   -s_{23}   &   c_{23}
                                       \end{array}
                                                     \right)
         \left(   \begin{array}{ccc}
                 c_{13}  &        0       &   s_{13} e^{-i \delta}         \\
                 0          &        1       &     0                                   \\
                 -s_{13} e^{ i \delta}          &   0   &   c_{13}
                                       \end{array}
                                                     \right)
         \left(   \begin{array}{ccc}
                  c_{12}   &   s_{12}   &   0       \\
                 -s_{12}   &   c_{12}   &   0       \\
                 0             &         0     &    1 
                                     \end{array}
                                                     \right)
          \left(   \begin{array}{ccc}
                       1      &      0                   &      0        \\
                      0       &     e^{ i \alpha}    &     0        \\
                       0      &      0                    &    e^{ i \beta}  
                                       \end{array}
                                                     \right),
  \label{bb}
\end{eqnarray}
where $s_{ij}$ and $ c_{ij}$ are sine and cosine of the lepton mixing angle $\theta_{ij}$,
   respectively.
The parameter $\delta$ is the CP-violating Dirac phase, and $ \alpha, \beta$ represent 
   the CP-violating Majorana phases.
Among nine parameters, $(m_1, m_2, m_3)$, $(\theta_{12}, \theta_{23}, \theta_{13})$,
   $\delta$, $(\beta, \alpha)$, the following five quantities are known by experiments:
   i.e., two mass squared differences $  \bigtriangleup m_{\odot}^2 = m_2^2 - m_1^2$,
   $  \bigtriangleup  m_{\rm A}^2  =  \vert m_3^2 - m_1^2 \vert
                           \approx \vert m_3^2 - m_2^2 \vert $,
  and three mixing angles $(\theta_{12}, \theta_{23}, \theta_{13})$.
These oscillation parameters are known to be within
\begin{eqnarray}
\bigtriangleup m_{\odot}^2 & = & (7.00 - 8.09) \times
                                   10^{-5} \, {\rm eV}^2,                  \nonumber  \\
    \bigtriangleup  m_{\rm A}^2  & = & (2.276 - 2.695) \times
                                   10^{-3} \, {\rm eV}^2,                  \nonumber  \\
   \sin^2 \theta_{12}  & = & (0.267 - 0.344),            \nonumber  \\
   \sin^2 \theta_{23}  & = & (0.342 - 0.667),            \nonumber  \\
   \sin^2 \theta_{13}  & = & (0.0156 - 0.0299),
      \label{bc}
\end{eqnarray}
at $3 \sigma$ \cite{rf:Gon}, respectively.
We therefore have the relations
   $  \bigtriangleup m_{\odot}^2 <<  \bigtriangleup  m_{\rm A}^2 $
   and $ \vert U_{e 1} \vert^2  >   \vert U_{e 2} \vert^2 >>   \vert U_{e 3} \vert^2  $.
   
The effective Majorana mass $\vert m_{ee} \vert $ can be written as 
\begin{equation}
    \vert m_{ee} \vert
   =   \left\vert \, \vert m_{ee}^{(1)} \vert +  \vert m_{ee}^{(2)} \vert e^{2 i \alpha}
    +  \vert m_{ee}^{(3)} \vert e^{2 i \beta} \, \right\vert,
  \label{bd}
\end{equation}
where
\begin{eqnarray}
     \vert m_{ee}^{(1)} \vert  &\equiv& m_1 \vert U_{e 1} \vert^2 = m_1 \, c_{1 2}^2 c_{1 3}^2,             \nonumber  \\
     \vert m_{ee}^{(2)} \vert  &\equiv& m_2 \vert U_{e 2} \vert^2 = m_2 \, s_{1 2}^2 c_{1 3}^2,             \nonumber  \\
     \vert m_{ee}^{(3)} \vert  &\equiv& m_3 \vert U_{e 3} \vert^2 = m_3 \, s_{1 3}^2,
  \label{be}
\end{eqnarray}
and
\begin{equation}
    \vert U_{e 1} \vert^2 +  \vert U_{e 2} \vert^2 +  \vert U_{e 3} \vert^2 =1.
\end{equation}
When obtaining the isocontours of the effective Majorana mass $ \vert m_{ee} \vert $
   in the $ \beta \alpha-$plane, the order of three $  \vert m_{ee}^{(i)} \vert $
   according to size becomes an important issue, as will be discussed in detail
   in section 3 and 4.
The effective Majorana mass $ \vert m_{ee} \vert $ is invariant under 
   $(\beta, \alpha)  \rightarrow   (-\beta, -\alpha) $ as well as under
   $  \beta \rightarrow \beta + n \pi, (n \in {\bf Z}) $ or 
   $  \alpha \rightarrow \alpha + m \pi, (m \in {\bf Z}) $.
The $ \vert m_{ee} \vert $ depends on seven parameters:
   mixing angles $(\theta_{12}, \theta_{13})$, neutrino mass $(m_1, m_2, m_3)$, 
   and CP-violating Majorana phases $(\beta, \alpha)$.
For these seven parameters,
   $\theta_{12}$, $ \theta_{13}$,
   $ \bigtriangleup m_{\odot}^2 $, and
   $ \bigtriangleup  m_{\rm A}^2 $
   are measured by experiments \cite{rf:Gon},
   while the Majorana phases $\beta, \alpha$ (if Majorana particles) 
   and the absolute neutrino mass scale are not measured.
Recently, an upper limit for the sum of the three light neutrino mass has been reported
   by Planck measurements of the cosmic microwave background \cite{rf:Ade,rf:HabTak},
\begin{equation}
   m_1 +m_2 +m_3 < 0.23 \, {\rm eV}  \hskip0.5cm (Planck + {\rm WP+high L+BAO }).
\end{equation}

In this paper, we regard the four parameters, $\theta_{12}, \theta_{13}$,
   $ \bigtriangleup m_{\odot}^2 $, and $ \bigtriangleup  m_{\rm A}^2 $ as given quantities.
The absolute neutrino mass scale is treated as a free parameter.
Then, the effective Majorana mass $ \vert m_{ee} \vert $ can be considered to be a function
   of two variables $\beta$ and $\alpha$ when one sets the size of the absolute
   neutrino mass scale.
There remains two possibilities with respect to the order of $m_1$, $ m_2$, and $ m_3$,
   that is the normal mass ordering $(m_3 >m_2>m_1)$ and inverted mass ordering
   $(m_2 >m_1>m_3) $.
Characteristic features of each mass ordering are discussed in the following.

In the normal mass ordering $(m_3 >m_2>m_1)$, we take the lightest neutrino mass
   $m_1$ as the absolute neutrino mass scale, and regard it as a free parameter,
\begin{eqnarray}
   m_2 &=& \sqrt{ m_1^2 + \bigtriangleup m_{\odot}^2 },                  \nonumber  \\
   m_3 &=&  \sqrt{ m_1^2 +   \bigtriangleup  m_{\rm A}^2 }.
  \label{bf}
\end{eqnarray}
The $ \vert m_{ee}^{(i)} \vert $ defined in Eqs.(\ref{be}) are represented by use of the
   lightest neutrino mass $m_1$,
\begin{eqnarray}
     \vert m_{ee}^{(1)} \vert  & = &  m_1 \, \vert U_{e 1} \vert^2,             \nonumber  \\
     \vert m_{ee}^{(2)} \vert  &=& 
          \sqrt{ m_1^2 + \bigtriangleup m_{\odot}^2 } ~  \vert U_{e 2} \vert^2,      \nonumber  \\
     \vert m_{ee}^{(3)} \vert  &=& 
                   \sqrt{ m_1^2 +   \bigtriangleup  m_{\rm A}^2 } ~  \vert U_{e 3} \vert^2.
  \label{bg}
\end{eqnarray}
Setting the value of the lightest neutrino mass $m_1$, we can regard the effective  
   Majorana mass $ \vert m_{ee} \vert $ as the function of $\beta$ and $\alpha$.
Now let us consider the order of $  \vert m_{ee}^{(1)} \vert $, $  \vert m_{ee}^{(2)} \vert $,
   and $  \vert m_{ee}^{(3)} \vert $ according to size in the normal mass ordering case.
For an arbitrary value of the lightest neutrino mass $m_1$, the following inequality holds,
\begin{equation}
     \frac{  \vert m_{ee}^{(3)} \vert }{ \vert m_{ee}^{(2)} \vert }
     = \frac{  \vert U_{e 3} \vert^2 }{ \vert U_{e 2} \vert^2 }  {m_3 \over m_2}
     \le   \frac{  \vert U_{e 3} \vert^2 }{  \vert U_{e 2} \vert^2 } \,
       \sqrt{ \frac{ \bigtriangleup  m_{\rm A}^2 }{  \bigtriangleup m_{\odot}^2 } }.
   \label{bh}
\end{equation}
By use of the experimental values \cite{rf:Gon}, Eq.(\ref{bc}), one has the relation
   $   \vert m_{ee}^{(3)} \vert  <  \vert m_{ee}^{(2)} \vert $ at  $ 3 \sigma$.
Taking account of this, we assume in this paper that the inequality,
\begin{equation}
    \vert m_{ee}^{(3)} \vert  <  \vert m_{ee}^{(2)} \vert,
  \label{bi}
\end{equation}
is satisfied for an arbitrary value of $m_1$.
We examine the order of $  \vert m_{ee}^{(1)} \vert $, $  \vert m_{ee}^{(2)} \vert $,
   and $  \vert m_{ee}^{(3)} \vert $ according to size when the value of the
   lightest neutrino mass $m_1$ is varied.
When taking the $ m_1 \rightarrow 0 $ limit, we have
\begin{equation}
     \vert m_{ee}^{(1)} \vert  <   \vert m_{ee}^{(3)} \vert.
  \label{bj}
\end{equation}
On the other hand, for $ m_1^2 >>  \bigtriangleup m_{\odot}^2 $, it follows
\begin{equation}
     \vert m_{ee}^{(2)} \vert  <   \vert m_{ee}^{(1)} \vert,
  \label{bk}
\end{equation}
because of $ \vert U_{e 2} \vert^2 <  \vert U_{e 1} \vert^2 $ and
   $  m_2  =   m_1 \sqrt{ 1 +  \bigtriangleup m_{\odot}^2 / m_1^2  }
            \approx m_1 $.
Therefore, the $  \vert m_{ee}^{(1)} \vert $ which is proportional to $m_1$ changes from
   zero to the maximum of $  \vert m_{ee}^{(i)} \vert \, (i=1,2,3) $ when the value of the
   lightest neutrino mass $m_1$ is varied.
For a concrete illustration, we show in Fig.1 the relation of the lightest neutrino mass
   $m_1$ and $  \vert m_{ee}^{(i)} \vert \, (i=1,2,3) $ when the values of parameters are
   set as $ \bigtriangleup m_{\odot}^2 = 7.55 \times 10^{-5} {\rm eV}^2 $,
             $ \bigtriangleup m_{\rm A}^2 = 2.486 \times 10^{-3} {\rm eV}^2 $,
             $ \sin ^2 \theta_{12} = 0.306 $, and $ \sin ^2 \theta_{13} = 0.0228 $.  
%
%
%
\begin{figure}
     \centering
     \includegraphics[height=7cm]{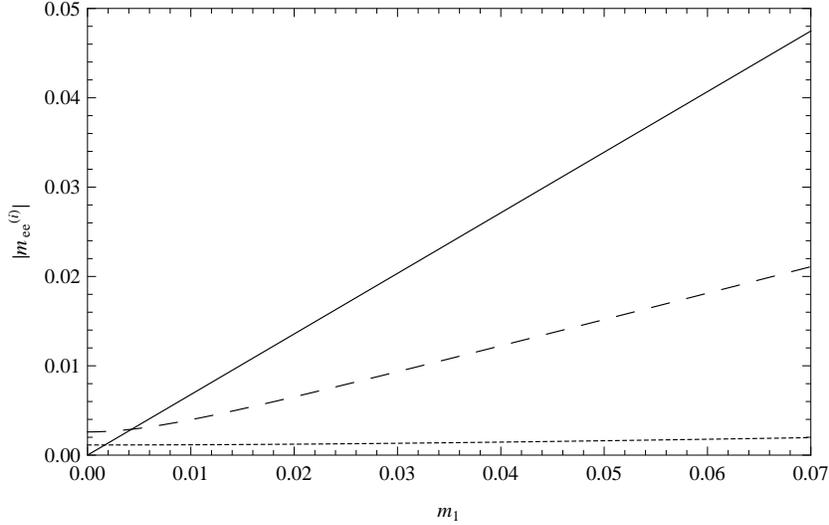}
    \caption{The value (in {\rm eV}) of $ \vert m_{ee}^{(1)} \vert $ (Solid), 
                 $ \vert m_{ee}^{(2)} \vert $ (long-dashed), and
                 $ \vert m_{ee}^{(3)} \vert $ (dotted) as a function of the
                 lightest neutrino mass $m_1$ (in {\rm eV}), respectively, for the normal
                 mass ordering case.}
\label{fig:1}
\end{figure}%
\noindent
After section 3, we do not specify the values of 
   $  \bigtriangleup m_{\odot}^2, \,  \bigtriangleup m_{\rm A}^2 $,
   $  \sin ^2 \theta_{12}$, and $ \sin ^2 \theta_{13}$.
For the lightest neutrino mass $m_1$ which is regarded as a free parameter, we do not
   give the value of $m_1$ but specify the size of $  \vert m_{ee}^{(1)} \vert $
   which is proportional to $m_1$.

In the inverted mass ordering ($m_2 >m_1>m_3$), we take the lightest neutrino mass
   $m_3$ as the absolute neutrino mass scale, and regard it as a free parameter,
\begin{eqnarray}
   m_2 &=& \sqrt{ m_3^2 + \bigtriangleup m_{\odot}^2 +  \bigtriangleup  m_{\rm A}^2 },                  \nonumber  \\
   m_1 &=&  \sqrt{ m_3^2 +   \bigtriangleup  m_{\rm A}^2 }.
  \label{bl}
\end{eqnarray}
The $ \vert m_{ee}^{(i)} \vert $ defined in Eq.(\ref{be}) are represented by use of the
    lightest neutrino mass $m_3$,
\begin{eqnarray}
     \vert m_{ee}^{(1)} \vert  &=&
          \sqrt{ m_3^2 +   \bigtriangleup  m_{\rm A}^2 } ~  \vert U_{e 1} \vert^2,     \nonumber  \\
     \vert m_{ee}^{(2)} \vert  &=& 
          \sqrt{ m_3^2 + \bigtriangleup m_{\odot}^2 +  \bigtriangleup  m_{\rm A}^2 } ~
                \vert U_{e 2} \vert^2,                                                 \nonumber  \\
     \vert m_{ee}^{(3)} \vert  & = & m_3 \,  \vert U_{e 3} \vert^2.
  \label{bm}
\end{eqnarray}
Setting the value of the lightest neutrino mass $m_3$, we can regard the effective
  Majorana mass $  \vert m_{ee} \vert $ as a function of $\beta$ and $\alpha$.
For an arbitrary value of the lightest neutrino mass $m_3$, we have
\begin{equation}
   \vert m_{ee}^{(1)} \vert >   \vert m_{ee}^{(2)} \vert  >>  \vert m_{ee}^{(3)} \vert,
  \label{bn}
\end{equation}
in the inverted mass ordering because of
   $ \vert U_{e 1} \vert^2  >   \vert U_{e 2} \vert^2 >>   \vert U_{e 3} \vert^2 $
   and $ m_1 \approx m_2 $.
The later relation, $ m_1 \approx m_2 $, comes from the following inequality,
\begin{equation}
  1 < { m_2 \over m_1 }
     = \sqrt{ 1 + \frac{  \bigtriangleup  m_{\odot}^2  }{  m_3^2 +   \bigtriangleup  m_{\rm A}^2 } }
     <   \sqrt{ 1 + \frac{  \bigtriangleup  m_{\odot}^2  }{  \bigtriangleup  m_{\rm A}^2 } }
     \approx 1.018.
  \label{bo}
\end{equation}
For a concrete illustration, we show in Fig.2 the relation of the lightest neutrino mass
   $m_3$ and $  \vert m_{ee}^{(i)} \vert \, (i=1,2,3) $ when the values of
   $  \bigtriangleup m_{\odot}^2, \,  \bigtriangleup m_{\rm A}^2 $,
   $  \sin ^2 \theta_{12}$, and $ \sin ^2 \theta_{13}$
   are the same with those of Fig.1.
%
%
%
%
\begin{figure}
     \centering
     \includegraphics[height=7cm]{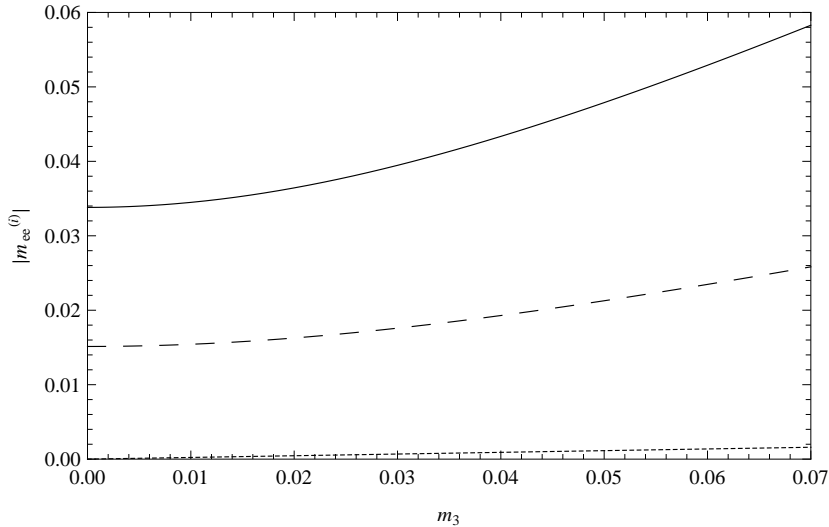}
    \caption{The value (in {\rm eV}) of $ \vert m_{ee}^{(1)} \vert $ (Solid), 
                 $ \vert m_{ee}^{(2)} \vert $ (long-dashed), and
                 $ \vert m_{ee}^{(3)} \vert $ (dotted) as a function of the
                 lightest neutrino mass $m_3$ (in {\rm eV}), respectively,
                 for the inverted mass ordering case.}
\label{fig:2}
\end{figure}%
\noindent                                                                                
As discussed before, we do not specify the values of
   $  \bigtriangleup m_{\odot}^2, \,  \bigtriangleup m_{\rm A}^2 $,
   $  \sin ^2 \theta_{12}$, and $ \sin ^2 \theta_{13}$
   after section 3.
%
%
%
\section{Majorana phases for the normal mass ordering}
One can expect that the $ 0 \, \nu \beta \beta$ experiment will bring the information on
   the Majorane phases $\beta$ and $\alpha$ in the future.
In this chapter, we derive the equation representing the isocontour of the effective
   Majorana mass $ \vert m_{ee} \vert $ in the $ \beta \alpha-$plane in the normal
   mass ordering case ($ m_3 >m_2>m_1 $).
As discussed in section 2, the lightest neutrino mass $m_1$ is treated as a free
   parameter, and when we set the value of $m_1$, the effective Majorana mass
   $ \vert m_{ee} \vert $ is considered as a function of $\beta$ and $\alpha$.
If the value of $ \vert m_{ee} \vert $ is determined, the constraint on $\beta$ and $\alpha$
   (isocontour of $ \vert m_{ee} \vert $ in the $ \beta \alpha-$plane) is obtained;
   however, one can not obtain both values of $\beta$ and $\alpha$ simultaneously.
We first obtain the isocontour of $ \vert m_{ee} \vert $ around the point of maximum
   $ \vert m_{ee} \vert $ in the $ \beta \alpha-$plane, and next those around the point of
   minimum $ \vert m_{ee} \vert $.
The point of maximum or minimum $ \vert m_{ee} \vert $ will be found in the region,
\begin{equation}
  - {\pi \over 2}   <  \beta \le  {\pi \over 2},   \hskip0.8cm
  - {\pi \over 2}   \le  \alpha  <  {\pi \over 2}.
  \label{cx}
\end{equation}
In the last subsection, we look for the isocontour except around the point of
   maximum or minimum $ \vert m_{ee} \vert $.
All these calculations are done on the assumption that the relation
   $ \vert m_{ee}^{(3)} \vert  <  \vert m_{ee}^{(2)} \vert $ holds as discussed in section 2.
\subsection{Around maximum value of $  \vert m_{ee} \vert  $.}
For an arbitrary value of the lightest neutrino mass $m_1$,  the effective Majorana mass
   $ \vert m_{ee} \vert $ takes a maximum value,
\begin{equation}
    \vert m_{ee} \vert_{\rm max}
     =  \vert m_{ee}^{(1)} \vert +  \vert m_{ee}^{(2)} \vert
          +  \vert m_{ee}^{(3)} \vert,
  \label{caa}
\end{equation}
at the point $(\beta, \alpha) = (0,0) $ in the $ \beta \alpha-$plane.
In order to obtain the isocontour around $(\beta, \alpha) = (0,0) $, we expand 
   $ \vert m_{ee} \vert^2 $ around this point.
Since the dependence of $ \vert m_{ee} \vert^2 $ on $\beta$ and $\alpha$ is
\begin{eqnarray}
    \vert m_{ee} \vert^2
     & = &  \vert m_{ee}^{(1)} \vert^2 +  \vert m_{ee}^{(2)} \vert^2 + \vert m_{ee}^{(3)} \vert^2
                + 2  \vert m_{ee}^{(1)} \vert  \vert m_{ee}^{(2)} \vert  \cos 2 \alpha
             + 2  \vert m_{ee}^{(1)} \vert  \vert m_{ee}^{(3)} \vert  \cos 2 \beta
                                                                                                                    \nonumber  \\
     & &  + 2  \vert m_{ee}^{(2)} \vert  \vert m_{ee}^{(3)} \vert  \cos 2 \alpha  \cos 2 \beta
            + 2  \vert m_{ee}^{(2)} \vert  \vert m_{ee}^{(3)} \vert  \sin 2 \alpha  \sin 2 \beta,
  \label{cab}
\end{eqnarray}
there appears no odd order term on the two variables $\beta$ and $\alpha$ in
   that expansion.
If one includes the terms up to second order on the two variables $\beta$ and $\alpha$
   in the expansion and disregards the higher orders, it becomes
\begin{eqnarray}
   \vert m_{ee} \vert^2
    & \approx &  \vert m_{ee} \vert_{\rm max}^2
                   - 4 \beta^2 \biggl\{ (   \vert m_{ee}^{(1)} \vert + \vert m_{ee}^{(2)} \vert )
                                     \vert m_{ee}^{(3)} \vert \biggr\}                             \nonumber  \\
    &  &  ~~   - 4 \alpha^2 \biggl\{ (   \vert m_{ee}^{(1)} \vert + \vert m_{ee}^{(3)} \vert )
                                     \vert m_{ee}^{(2)} \vert \biggr\}
                     + 8 \beta  \alpha  \vert m_{ee}^{(2)} \vert   \vert m_{ee}^{(3)} \vert,
  \label{cac}
\end{eqnarray}
or
\begin{equation}
     \left(     \begin{array}{cc}
        \beta  &   \alpha
                 \end{array}
                                 \right)
    \left(     \begin{array}{cc}
        a  &   b         \\
        b  &  c
                 \end{array}
                                 \right)
   \left(     \begin{array}{c}
         \beta            \\
         \alpha
                 \end{array}
                                 \right)
      \approx  J,
     \label{cad}
\end{equation}
where
\begin{eqnarray}
   a   & \equiv &   (   \vert m_{ee}^{(1)} \vert + \vert m_{ee}^{(2)} \vert )
                                 \vert m_{ee}^{(3)} \vert   > 0,                                      \nonumber  \\
   b   & \equiv &   -   \vert m_{ee}^{(2)} \vert   \vert m_{ee}^{(3)} \vert  \hskip1.3cm   < 0,      \nonumber  \\
   c   & \equiv &   (   \vert m_{ee}^{(1)} \vert + \vert m_{ee}^{(3)} \vert )
                                 \vert m_{ee}^{(2)} \vert   > 0,                                   \nonumber  \\
   J  & \equiv &    {1 \over 4}  \biggl\{   \vert m_{ee} \vert_{\rm max}^2 -  \vert m_{ee} \vert^2
                                                                                                                   \biggr\} > 0.
  \label{cae}
\end{eqnarray}

A real symmetric matrix can be diagonalized by a rotational matrix,
\begin{equation}
     \left(     \begin{array}{cc}
        \cos \theta   &   \sin \theta         \\
        - \sin \theta  &  \cos \theta
                 \end{array}
                                                     \right)
      \left(     \begin{array}{cc}
        a  &   b         \\
        b  &  c
                 \end{array}
                                                   \right)
      \left(     \begin{array}{cc}
        \cos \theta   &   - \sin \theta         \\
        \sin \theta  &  \cos \theta
                 \end{array}
                                                     \right)
   =  \left(     \begin{array}{cc}
        \lambda_{-}  &   0         \\
        0                   &    \lambda_{+}
                 \end{array}
                                                    \right),
  \label{caf}
\end{equation}
where
\begin{eqnarray}
   \tan \theta & =& \frac{ c - a - \sqrt{ (a - c)^2 + 4 b^2 } }{2 b },   \hskip1.5cm  (b \ne 0),
                                                                                                             \nonumber  \\
   \tan 2 \theta & =& \frac{ 2 b}{a-c},     \hskip4.4cm          (a-c \ne 0),     \nonumber  \\
   \lambda_{\pm} &=& \frac{ (a+c) \pm \sqrt{ (a - c)^2 + 4 b^2 } }{2}.
  \label{cag}
\end{eqnarray}
If $ a c - b^2 > 0 $ and $ a+c > 0 $ are satisfied, one has
   $ \lambda_{+} > \lambda_{-} > 0 $.
   
Since $ a+c > 0 $ and $ a c - b^2 > 0 $ are satisfied from Eq.(\ref{cae}), a homogeneous
   polynomial of second order, Eq.(\ref{cad}), represents the ellipse in
   the $ \beta \alpha-$plane.
By diagonalization of the real symmetric matrix, Eq.(\ref{cad}) can be rewritten as
\begin{equation}
     \left(     \begin{array}{cc}
        \beta''  &   \alpha''
                 \end{array}
                                 \right)
    \left(     \begin{array}{cc}
        \lambda_{-}  &   0         \\
        0                   &    \lambda_{+}
                 \end{array}
                                 \right)
   \left(     \begin{array}{c}
         \beta''            \\
         \alpha''
                 \end{array}
                                 \right)
      \approx  J,
     \label{cah}
\end{equation}
where
\begin{equation}
    \left(     \begin{array}{c}
         \beta''            \\
         \alpha''
                 \end{array}
                                 \right)
     =
     \left(     \begin{array}{cc}
        \cos \theta   &   \sin \theta         \\
        - \sin \theta  &  \cos \theta
                 \end{array}
                                 \right)
     \left(     \begin{array}{c}
         \beta            \\
         \alpha
                 \end{array}
                                 \right),
  \label{cai}
\end{equation}
and
\begin{eqnarray}
   \tan \theta & =&  \frac{ - 1 }{ 2    \vert m_{ee}^{(2)} \vert   \vert m_{ee}^{(3)} \vert }
                         \biggl[    \vert m_{ee}^{(1)} \vert  (  \vert m_{ee}^{(2)} \vert  - \vert m_{ee}^{(3)} \vert  )    \biggr.                \nonumber  \\
     &    &  ~~   \left.   -  \sqrt{  \vert m_{ee}^{(1)} \vert^2 ( \vert m_{ee}^{(3)} \vert -  \vert m_{ee}^{(2)} \vert )^2
               + 4 (  \vert m_{ee}^{(2)} \vert \vert m_{ee}^{(3)} \vert )^2  } \,  \right] > 0,   \nonumber    \\ \tan 2 \theta & =&  \frac{2    \vert m_{ee}^{(2)} \vert   \vert m_{ee}^{(3)} \vert }
                          {   \vert m_{ee}^{(1)} \vert (   \vert m_{ee}^{(2)} \vert -  \vert m_{ee}^{(3)} \vert ) }>0,
                                                              \nonumber  \\
  \lambda_{\pm} &=&  {1 \over 2} \left[  \biggl\{
                   \vert m_{ee}^{(1)} \vert  \vert m_{ee}^{(3)} \vert  + 2  \vert m_{ee}^{(2)} \vert   \vert m_{ee}^{(3)} \vert
                     +   \vert m_{ee}^{(1)} \vert  \vert m_{ee}^{(2)} \vert  \biggr\}    \right.                \nonumber  \\
     &    &  ~~   \left.   \pm \sqrt{  \vert m_{ee}^{(1)} \vert^2 ( \vert m_{ee}^{(3)} \vert -  \vert m_{ee}^{(2)} \vert )^2
               + 4 (  \vert m_{ee}^{(2)} \vert \vert m_{ee}^{(3)} \vert )^2  } \,  \right].
  \label{caj}
\end{eqnarray}
As $ \tan \theta $ and $ \tan 2 \theta \, (= 2 \tan \theta / (1-\tan^2 \theta)) $ are
   both positive, we have $ 0 < \theta < {\pi / 4} $.
In the $ \beta \alpha-$plane, the isocontour around the point of maximum
   $ \vert m_{ee} \vert_{\rm max} $ becomes the ellipse represented by the following
   normal equation,
\begin{equation}
   \frac{ \beta''^2 }{ \left( \sqrt{ J \over \lambda_-  } \right)^2 }
   +  \frac{ \alpha''^2 }{ \left( \sqrt{ J \over \lambda_+  } \right)^2 } = 1,
   \label{cak}
\end{equation}
where we have neglected the fourth order of two variables $ \beta $ and $ \alpha $
   or higher orders.
The center of this ellipse is $(\beta, \alpha) = (0,0) $, and the major axis is
   $ 2 \sqrt{ J/ \lambda_{-} } $, the minor axis $ 2 \sqrt{ J/ \lambda_{+} } $.
The ratio of the minor axis to the major axis, $ \sqrt{  \lambda_{-} /  \lambda_{+} } $,
   depends on the lightest neutrino mass $m_1$, but not on $ \vert m_{ee} \vert $.
The direction of the $\beta''$ axis is produced by a counterclockwise rotation of the
   $\beta$ axis by the angle $\theta (>0)$ (see Fig.3).
%
%
%
%
\begin{figure}
     \centering
     \includegraphics[height=7cm]{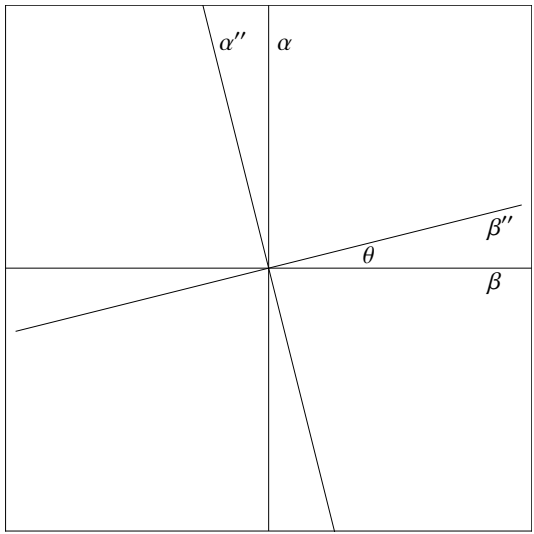}
    \caption{Relation between the $\beta''$ axis (the $\alpha''$ axis)
                    and  the $\beta$ axis (the $\alpha$ axis).}
\label{fig:3}
\end{figure}%
\noindent                                                                                   
The effective Majorana mass, Eq.(\ref{ab}), is invariant under
   $ \beta \rightarrow \beta + n \pi,
       ~  \alpha \rightarrow \alpha + m \pi, ( n,m \in {\bf Z} )$,
   and consequently the ellipse with the center $( n \pi, m \pi )$ is distributed in
   the $ \beta \alpha-$plane.
When the size of the major axis is a quarter of $ \pi$, $ 2 \sqrt{ J/ \lambda_{-} } = \pi / 4$,
   the effective Majorana mass becomes
\begin{equation}
   \vert m_{ee} \vert = \sqrt{   \vert m_{ee} \vert_{\rm max}^2 
            - \frac{\pi^2}{16} \lambda_{-} } \, .
   \label{cal}
\end{equation}

The ratio of the minor axis to the major axis $ \sqrt{  \lambda_{-} /  \lambda_{+} } $
   reflects the form of the ellipse,
   the major axis reflects the size of the ellipse,
   and the angle $ \theta $ determines the direction of the major axis of the ellipse.
Now, we investigate how $ \sqrt{  \lambda_{-} /  \lambda_{+} } $ and $ \theta $
   change when the value of the lightest neutrino mass $m_1$ is varied.
We also examine the relation between the effective Majorana mass $ \vert m_{ee} \vert $
   and the major axis of the ellipse which is the isocontour of $ \vert m_{ee} \vert $.
The expressions of $\lambda_{\pm}$ or $ \tan \theta $ are, however, too complicated
   to study these problems practically.
As discussed in section 2, the $  \vert m_{ee}^{(1)} \vert $ changes from zero to 
   the maximum of $  \vert m_{ee}^{(i)} \vert \, (i=1,2,3) $ when the value of 
   the lightest neutrino mass $m_1$ is varied.
When the value of $m_1$ (or $  \vert m_{ee}^{(1)} \vert = m_1  \vert U_{e 1} \vert^2 $)
   is in some regions, $\lambda_{\pm}$ or $ \tan \theta $ can be expressed simply,
   and we give three instances in the following.

The first case is that the value of the lightest neutrino mass $m_1$ is large enough to
   satisfy inequality,
\begin{equation}
   \vert m_{ee}^{(1)} \vert >   \vert m_{ee}^{(2)} \vert  >>  \vert m_{ee}^{(3)} \vert,
   ~~{\rm and}~~   m_1 \approx m_2.
  \label{cam}
\end{equation}
Owing to this inequality, the terms of order $O(\vert m_{ee}^{(3)} \vert^2)$ can be
   neglected,
\begin{eqnarray}
   \tan \theta  & \approx & \frac{  \vert m_{ee}^{(3)} \vert }{  \vert m_{ee}^{(1)} \vert } \,
                                            = \, { m_3 \over m_1 } \frac{ \vert U_{e 3} \vert^2 }{ \vert U_{e 1} \vert^2 } << 1,
                                                                       \nonumber  \\
   \lambda_{+}  & \approx &  \vert m_{ee}^{(1)} \vert  \vert m_{ee}^{(2)} \vert
                                               +   \vert m_{ee}^{(2)} \vert  \vert m_{ee}^{(3)} \vert,        \nonumber  \\
   \lambda_{-}  & \approx &  ( \vert m_{ee}^{(1)} \vert + \vert m_{ee}^{(2)} \vert )
                                                    \vert m_{ee}^{(3)} \vert.
   \label{can}
\end{eqnarray}
The ratio of the minor axis to the major axis can be written as
\begin{eqnarray}
   \sqrt{ \lambda_{-} \over \lambda_{+} }
   \approx \sqrt{ \frac{ (  \vert m_{ee}^{(1)} \vert +  \vert m_{ee}^{(2)} \vert ) }
                                {   \vert m_{ee}^{(1)} \vert   \vert m_{ee}^{(2)} \vert }   \vert m_{ee}^{(3)} \vert  } 
    \approx \sqrt{ m_3 \over m_1 }
       \sqrt{ \frac{ \vert U_{e 3} \vert^2 ( 1 -   \vert U_{e 3} \vert^2 ) }{ \vert U_{e 1} \vert^2  \vert U_{e 2} \vert^2 } },
  \label{cao}
\end{eqnarray}
where we have used $  m_1 \approx m_2 $.
In the region where the condition Eq.(\ref{cam}) is satisfied,
   both the angle $\theta$ and the ratio $ \sqrt{  \lambda_{-} /  \lambda_{+} } $ increase
   when the value of the the lightest neutrino mass $m_1$ is decreased.
When the size of the major axis of the ellipse is $ \pi / 4$, the effective Majorana mass
   Eq.(\ref{cal}) becomes
\begin{equation}
      \vert m_{ee} \vert  \approx   \vert m_{ee} \vert_{\rm max} 
                                           - {\pi^2 \over 32}   \vert m_{ee}^{(3)} \vert         
                =   \vert m_{ee}^{(1)} \vert + \vert m_{ee}^{(2)} \vert 
                                  + \left( 1- {\pi^2 \over 32} \right)  \vert m_{ee}^{(3)} \vert.
  \label{cap}
\end{equation}
In other words, when the effective Majorana mass takes the value Eq.(\ref{cap}),
   the corresponding isocontour is the ellipse represented by
\begin{equation}
   \frac{\beta''^2}{ \left(  \pi \over 8 \right)^2 }
   + \frac{\alpha''^2}
        { \left( { \pi \over 8 } \sqrt{  \frac{ (  \vert m_{ee}^{(1)} \vert +  \vert m_{ee}^{(2)} \vert ) }
                          {   \vert m_{ee}^{(1)} \vert   \vert m_{ee}^{(2)} \vert }   \vert m_{ee}^{(3)} \vert } 
                \,  \right)^2 }
   = 1.
     \label{caq}
\end{equation}

The second case is that the the lightest neutrino mass $m_1$ has smaller value than
   the first case and satisfies
\begin{equation}
   \vert m_{ee}^{(1)} \vert =  \vert m_{ee}^{(2)} \vert  +  \vert m_{ee}^{(3)} \vert.
  \label{car}
\end{equation}
In this case, we have
\begin{eqnarray}
   \tan \theta &=& \frac{  \vert m_{ee}^{(3)} \vert }{  \vert m_{ee}^{(2)} \vert } < 1,
                                                                                                         \nonumber \\
    \sqrt{ \lambda_{-} \over \lambda_{+} }
       &=&   \sqrt{   \frac{  2 \,  \vert m_{ee}^{(2)} \vert    \vert m_{ee}^{(3)} \vert }
                                    {  2 \,  \vert m_{ee}^{(2)} \vert    \vert m_{ee}^{(3)} \vert
                                     +  \vert m_{ee}^{(2)} \vert^2 +  \vert m_{ee}^{(3)} \vert^2 }   } < 1,   
  \label{cas}
\end{eqnarray}
and both the angle $\theta$ and the ratio $ \sqrt{  \lambda_{-} /  \lambda_{+} } $ are
   larger than the first case, respectively.
When the size of the major axis of the ellipse is $ \pi / 4$, the effective Majorana
   mass becomes
\begin{equation}
      \vert m_{ee} \vert  \approx   \vert m_{ee} \vert_{\rm max} - {\pi^2 \over 32} \,
                          \frac{  \vert m_{ee}^{(2)} \vert    \vert m_{ee}^{(3)} \vert }
                                    { (  \vert m_{ee}^{(2)} \vert +  \vert m_{ee}^{(3)} \vert ) }.
   \label{cat}
\end{equation}

The third case is that the lightest neutrino mass $m_1$ has much smaller value than
   the second case and satisfies
\begin{equation}
   \vert m_{ee}^{(2)} \vert >   \vert m_{ee}^{(3)} \vert  >>  \vert m_{ee}^{(1)} \vert.
  \label{cau}
\end{equation}
In this case, the terms of order $O(\vert m_{ee}^{(1)} \vert^2)$ can be neglected and
   one has
\begin{equation}
   \tan \theta   \approx
     1- \frac{  ( \vert m_{ee}^{(2)} \vert -   \vert m_{ee}^{(3)} \vert ) }{ 2 \,  \vert m_{ee}^{(2)} \vert  \vert m_{ee}^{(3)} \vert }
          \vert m_{ee}^{(1)} \vert,
  \label{cava}
\end{equation}
or
\begin{equation}
   \theta  \approx  { \pi \over 4 } - \frac{  ( \vert m_{ee}^{(2)} \vert -   \vert m_{ee}^{(3)} \vert ) }
                                                       { 4 \,  \vert m_{ee}^{(2)} \vert  \vert m_{ee}^{(3)} \vert }  \vert m_{ee}^{(1)} \vert   
          \approx  { \pi \over 4 },
  \label{cavb}
\end{equation}
and
\begin{eqnarray}
   \sqrt{ \lambda_{-} \over \lambda_{+} }
      & \approx &  \sqrt{   \frac{  ( \vert m_{ee}^{(2)} \vert +  \vert m_{ee}^{(3)} \vert ) }
                                    { 4  \vert m_{ee}^{(2)} \vert   \vert m_{ee}^{(3)} \vert }   \vert m_{ee}^{(1)} \vert  }.
       \label{cavc}
\end{eqnarray}
The angle $\theta$ is larger than the second case, while the ratio
   $ \sqrt{  \lambda_{-} /  \lambda_{+} } $ is much smaller than the second case.
In the region where the condition Eq.(\ref{cau}) is satisfied,
   the angle $\theta$ increases and the ratio
   $ \sqrt{  \lambda_{-} /  \lambda_{+} } $ decreases
   when the value of the the lightest neutrino mass $m_1$ is decreased.
In the $ m_1 \rightarrow 0 $ limit, we have
\begin{equation}
   \theta  \rightarrow   { \pi \over 4} ,    \hskip0.6cm
   \sqrt{ \lambda_{-} \over \lambda_{+} }   \rightarrow   0,  ~~~ (m_1 \rightarrow 0).
  \label{caw}
\end{equation}
When the size of the major axis of the ellipse is $ \pi / 4$, the effective Majorana 
   mass becomes
\begin{eqnarray}
      \vert m_{ee} \vert & \approx &  \vert m_{ee} \vert_{\rm max} 
                                           - {\pi^2 \over 64}   \vert m_{ee}^{(1)} \vert.
      \label{cax}                                       
\end{eqnarray}
\subsection{Around minimum value of $  \vert m_{ee} \vert  $.}
The position of the point $(\beta, \alpha)$ giving the minimum value of
   $  \vert m_{ee} \vert  $ shifts according to the value of the lightest neutrino mass
   $m_1$ (or $  \vert m_{ee}^{(1)} \vert $), because $  \vert m_{ee}^{(1)} \vert  $
   changes from zero to the maximum of $ \vert m_{ee}^{(i)} \vert \, (i=1,2,3) $ 
   when the value of $m_1$ is varied.
In studying the minimum value of $ \vert m_{ee} \vert $, it is convenient to divide
   $  \vert m_{ee}^{(1)} \vert  $ into the following three regions,
\begin{itemize}
  \item  region (A);     $ \hskip1cm   \vert m_{ee}^{(2)} \vert +  \vert m_{ee}^{(3)} \vert <  \vert m_{ee}^{(1)} \vert $,
  \item  region (B);   $ \hskip1cm   \vert m_{ee}^{(2)} \vert -   \vert m_{ee}^{(3)} \vert
                                    <  \vert m_{ee}^{(1)} \vert <  \vert m_{ee}^{(2)} \vert +  \vert m_{ee}^{(3)} \vert $,
   \item  region (C);    $ \hskip1cm   \vert m_{ee}^{(1)} \vert <   \vert m_{ee}^{(2)} \vert -   \vert m_{ee}^{(3)} \vert $.
\end{itemize}
The minimum value of $ \vert m_{ee} \vert $ in each of the three regions
   becomes \cite{rf:NunTevFun}
\begin{itemize}
  \item  region (A);  $ \hskip1cm     \vert m_{ee} \vert_{\rm min}
                                =  \vert m_{ee}^{(1)} \vert -  \vert m_{ee}^{(2)} \vert
                                    -  \vert m_{ee}^{(3)} \vert,  $
  \item  region (B);   $ \hskip1cm     \vert m_{ee} \vert_{\rm min}
                                =  0,  $
  \item  region (C);    $ \hskip1cm     \vert m_{ee} \vert_{\rm min}
                                =  \vert m_{ee}^{(2)} \vert -  \vert m_{ee}^{(3)} \vert
                                    -  \vert m_{ee}^{(1)} \vert.  $
\end{itemize}
In this subsection, we obtain the isocontour of $ \vert m_{ee} \vert $ in
   the $ \beta \alpha-$plane around the point of the minimum $ \vert m_{ee} \vert $
   in each of the three regions (A), (B), and (C).
Moreover, the characteristic features of the isocontour on the boundary between the
   region (A) and (B), or between (B) and (C), will be discussed.
\subsubsection{ The region (A) : 
    $  \vert m_{ee}^{(2)} \vert +  \vert m_{ee}^{(3)} \vert <  \vert m_{ee}^{(1)} \vert. $}
In the region (A), the effective Majorana mass has a minimum value,
   $ \vert m_{ee} \vert_{\rm min} $,
\begin{equation}
    \vert m_{ee} \vert_{\rm min}
     =  \vert m_{ee}^{(1)} \vert -  \vert m_{ee}^{(2)} \vert
          -  \vert m_{ee}^{(3)} \vert,
  \label{cbaa}
\end{equation}
at the point $(\beta, \alpha) = ({ \pi / 2 },  - { \pi / 2 })  $.
We expand $ \vert m_{ee} \vert^2 $ around this point to obtain the isocontour around
   the point.
In a similar way to the discussion in section3.1, there appears no odd order term on
   the two variables $(\beta - \pi / 2)$ and $ (\alpha + \pi / 2)$ in this expansion
   around $ (\beta, \alpha) = ({ \pi / 2 },  - { \pi / 2 })  $.
Up to the second order, the expansion results
\begin{eqnarray}
   \vert m_{ee} \vert^2
    & \approx &  \vert m_{ee} \vert_{\rm min}^2
                   + 4 ( \beta - { \pi \over 2 })^2 \biggl\{ (   \vert m_{ee}^{(1)} \vert - \vert m_{ee}^{(2)} \vert )
                                     \vert m_{ee}^{(3)} \vert \biggr\}                             \nonumber  \\
    &  & ~~  + 4 ( \alpha + { \pi \over 2 })^2 \biggl\{ (   \vert m_{ee}^{(1)} \vert - \vert m_{ee}^{(3)} \vert )
                                     \vert m_{ee}^{(2)} \vert \biggr\}                               \nonumber  \\
    &  & ~~    + 8 ( \beta - { \pi \over 2 }) ( \alpha + { \pi \over 2 })  \vert m_{ee}^{(2)} \vert   \vert m_{ee}^{(3)} \vert,
  \label{cbab}
\end{eqnarray}
or
\begin{equation}
     \left(   \begin{array}{cc}
                  \beta'   &    \alpha'
                 \end{array}
                                 \right)
    \left(     \begin{array}{cc}
        a  &   b         \\
        b  &   c
                 \end{array}
                                 \right)
   \left(     \begin{array}{c}
                  \beta'           \\
                  \alpha'
                 \end{array}
                                 \right)
      \approx  K,
     \label{cbac}
\end{equation}
where
\begin{equation}
   \left(   \begin{array}{c}
                  \beta'          \\
                  \alpha' 
              \end{array}
                                 \right)
     =   \left(     \begin{array}{c}
         ( \beta - { \pi \over 2 })            \\
         ( \alpha + { \pi \over 2 })
                 \end{array}
                                 \right),
  \label{cbad}
\end{equation}
\begin{eqnarray}
   a   & = &   (   \vert m_{ee}^{(1)} \vert - \vert m_{ee}^{(2)} \vert )
                                 \vert m_{ee}^{(3)} \vert   > 0,                                      \nonumber  \\
   b   & = &      \vert m_{ee}^{(2)} \vert   \vert m_{ee}^{(3)} \vert  \hskip1.3cm   > 0,      \nonumber  \\
   c   & = &   (   \vert m_{ee}^{(1)} \vert - \vert m_{ee}^{(3)} \vert )
                                 \vert m_{ee}^{(2)} \vert   > 0,                                   \nonumber  \\
   K  & = &    {1 \over 4}  \biggl\{   \vert m_{ee} \vert^2 - \vert m_{ee} \vert_{\rm min}^2
                                                                                                                   \biggr\} > 0,
  \label{cbae}
\end{eqnarray}
and one has $  a c - b^2 > 0 $.
By diagonalizing the real symmetric matrix as discussed in section 3.1, we can rewrite
   Eq.(\ref{cbac}) as
\begin{equation}
     \left(  \begin{array}{cc}
                \beta''   &    \alpha''
              \end{array}
                                 \right)
    \left(     \begin{array}{cc}
        \lambda_{-}  &   0         \\
        0                   &    \lambda_{+}
                 \end{array}
                                 \right)
   \left(   \begin{array}{c}
                \beta''          \\
                \alpha''
             \end{array}
                                 \right)
      \approx  K,
     \label{cbaf}
\end{equation}
where
\begin{equation}
    \left(   \begin{array}{c}
                \beta''          \\
                \alpha''
             \end{array}
                                 \right)
  =  \left(     \begin{array}{cc}
        \cos \theta   &   \sin \theta         \\
        - \sin \theta  &  \cos \theta
                 \end{array}
                                 \right)
     \left(   \begin{array}{c}
                   \beta'           \\
                   \alpha'
               \end{array}
                                 \right)
  =
     \left(     \begin{array}{cc}
        \cos \theta   &   \sin \theta         \\
        - \sin \theta  &  \cos \theta
                 \end{array}
                                 \right)
     \left(     \begin{array}{c}
         ( \beta - { \pi \over 2 })            \\
         ( \alpha + { \pi \over 2 })
                 \end{array}
                                 \right),
  \label{cbag}
\end{equation}
and
\begin{eqnarray}
   \tan \theta & =&  \frac{ + 1 }{ 2    \vert m_{ee}^{(2)} \vert   \vert m_{ee}^{(3)} \vert }
                         \biggl[    \vert m_{ee}^{(1)} \vert  (  \vert m_{ee}^{(2)} \vert  - \vert m_{ee}^{(3)} \vert  )    \biggr.                \nonumber  \\
     &    &  ~~   \left.   -  \sqrt{  \vert m_{ee}^{(1)} \vert^2 ( \vert m_{ee}^{(3)} \vert -  \vert m_{ee}^{(2)} \vert )^2
               + 4 (  \vert m_{ee}^{(2)} \vert \vert m_{ee}^{(3)} \vert )^2  } \,  \right]  <  0,   \nonumber    \\ 
    \tan 2 \theta & =&  - \frac{2    \vert m_{ee}^{(2)} \vert   \vert m_{ee}^{(3)} \vert }
                          {   \vert m_{ee}^{(1)} \vert (   \vert m_{ee}^{(2)} \vert -  \vert m_{ee}^{(3)} \vert ) } < 0,
                                                              \nonumber  \\
  \lambda_{\pm} &=&  {1 \over 2} \left[  \biggl\{
                   \vert m_{ee}^{(1)} \vert  \vert m_{ee}^{(3)} \vert  - 2  \vert m_{ee}^{(2)} \vert   \vert m_{ee}^{(3)} \vert
                     +   \vert m_{ee}^{(1)} \vert  \vert m_{ee}^{(2)} \vert  \biggr\}    \right.                \nonumber  \\
     &    &  ~~   \left.   \pm \sqrt{  \vert m_{ee}^{(1)} \vert^2 ( \vert m_{ee}^{(3)} \vert -  \vert m_{ee}^{(2)} \vert )^2
               + 4 (  \vert m_{ee}^{(2)} \vert \vert m_{ee}^{(3)} \vert )^2  } \,  \right].
  \label{cbah}
\end{eqnarray}
Because of $ \tan \theta < 0 $ and $ \tan 2 \theta < 0 $, one has
   $ - {\pi / 4} < \theta <  0 $.
In the region (A),
   $  \vert m_{ee}^{(2)} \vert +  \vert m_{ee}^{(3)} \vert <  \vert m_{ee}^{(1)} \vert $,
   the isocontour around the point of the minimum
   $ \vert m_{ee} \vert_{\rm min} = \vert m_{ee}^{(1)} \vert -  \vert m_{ee}^{(2)} \vert
                                    -  \vert m_{ee}^{(3)} \vert $
  becomes the ellipse represented by the following equation,
\begin{equation}
   \frac{ \beta''^2 }{ \left( \sqrt{ K \over \lambda_-  } \right)^2 }
   +  \frac{ \alpha''^2 }{ \left( \sqrt{ K \over \lambda_+  } \right)^2 } = 1,
   \label{cbai}
\end{equation}
where the fourth order of two variables $ (\beta - \pi / 2)$ and$ (\alpha + \pi / 2)$
   or higher orders have been neglected.
This ellipse has the center at $ (\beta, \alpha) =  ({ \pi / 2 },  - { \pi / 2}) $,
   and the major axis, $ 2 \sqrt{ K/ \lambda_{-} } $, the minor axis,
   $ 2 \sqrt{ K/ \lambda_{+} } $.
The direction of the $\beta''$ axis is produced by a clockwise rotation of the $\beta$
   axis by the angle $ \vert \theta \vert (>0)$.
From the invariance of $ \vert m_{ee} \vert $, Eq.(\ref{ab}), the ellipse with the center
   $ ({\pi / 2} + n \pi, - {\pi / 2} + m \pi)$ is distributed in the $ \beta \alpha-$plane.
When the size of the major axis is a quarter of $ \pi$, $ 2 \sqrt{ K/ \lambda_{-} } = \pi / 4$,
   the effective Majorana mass becomes
\begin{equation}
   \vert m_{ee} \vert = \sqrt{   \vert m_{ee} \vert_{\rm min}^2 
            + \frac{\pi^2}{16} \lambda_{-} } \, .
   \label{cbaix}
\end{equation}

As considered in section 3.1, we investigate how the ratio of the minor axis to the major
   axis, $  \sqrt{  \lambda_{-} /  \lambda_{+} }  $, reflecting the form of the ellipse, and
   the angle $\theta$ determining the direction of the major axis of the ellipse change
   when the value of the lightest neutrino mass $m_1$ is varied.
Although the expressions of $\lambda_{\pm}$ or $ \tan \theta $ in the region (A) are
   very complicated, these expressions are simplified when the value of $m_1$
   (or $  \vert m_{ee}^{(1)} \vert = m_1  \vert U_{e 1} \vert^2 $) satisfies the following,
\begin{equation}
   \vert m_{ee}^{(1)} \vert >   \vert m_{ee}^{(2)} \vert  >>  \vert m_{ee}^{(3)} \vert,
   ~~  m_1 \approx m_2.
  \label{cbaj}
\end{equation}
Neglecting the order $O(\vert m_{ee}^{(3)} \vert^2)$, we have
\begin{eqnarray}
   \theta  & \approx & - \frac{  \vert m_{ee}^{(3)} \vert }{  \vert m_{ee}^{(1)} \vert } \,
                                            = \, -{ m_3 \over m_1 } \frac{ \vert U_{e 3} \vert^2 }{ \vert U_{e 1} \vert^2 } < 0,  \hskip0.5cm   \vert \theta \vert << 1,                                \nonumber  \\
    \sqrt{ \lambda_{-} \over \lambda_{+} }
   & \approx &  \sqrt{ \frac{ (  \vert m_{ee}^{(1)} \vert -  \vert m_{ee}^{(2)} \vert ) }
                                {   \vert m_{ee}^{(1)} \vert   \vert m_{ee}^{(2)} \vert }   \vert m_{ee}^{(3)} \vert  } 
   \approx \sqrt{ m_3 \over m_1 }
       \sqrt{ \frac{ \vert U_{e 3} \vert^2 ( \vert U_{e 1} \vert^2 - \vert U_{e 2} \vert^2 ) }{ \vert U_{e 1} \vert^2  \vert U_{e 2} \vert^2 } }.
     \label{cbak}
\end{eqnarray}
In the region where the condition Eq.(\ref{cbaj}) is satisfied,
   both the absolute value of the angle $ \vert \theta \vert$ and the ratio
   $ \sqrt{  \lambda_{-} /  \lambda_{+} } $ increases
   when the value of the lightest neutrino mass $m_1$ is decreased.
When the size of the major axis of the ellipse is $ \pi / 4$, 
   the effective Majorana mass Eq.(\ref{cbaix}) becomes
\begin{equation}
      \vert m_{ee} \vert
           \approx   \vert m_{ee} \vert_{\rm min} 
                      + {\pi^2 \over 32}   \vert m_{ee}^{(3)} \vert
            =  \vert m_{ee}^{(1)} \vert - \vert m_{ee}^{(2)} \vert 
                   - \left( 1- {\pi^2 \over 32} \right)  \vert m_{ee}^{(3)} \vert.
  \label{cbaiy}
\end{equation}
\subsubsection{ On the boundary between the region (A) and region (B). }
Let us focus on the boundary between the region (A) and region (B),
\begin{equation}
   \vert m_{ee}^{(1)} \vert =  \vert m_{ee}^{(2)} \vert  +  \vert m_{ee}^{(3)} \vert.
  \label{cbba}
\end{equation}
On this boundary, the effective Majorana mass has a minimum value
   $ \vert m_{ee} \vert_{\rm min} =0 $ at
   $ (\beta, \alpha) =  ({ \pi / 2 },  - { \pi / 2 }) $.
When the $  \vert m_{ee} \vert^2  $ is expanded around the point 
   $ (\beta, \alpha) = ({ \pi / 2 },  - { \pi / 2 })  $, there appears no odd order term
   on the two variables $  (\beta - \pi / 2) $ and $ (\alpha + \pi / 2)$ in the expansion.
The expansion up to the second order takes the same form with Eq.(\ref{cbab}).
Because of $c-a > 0$ and $ ac-b^2 = 0$ on this boundary, one has
\begin{equation}
     \tan \theta  =   - \frac{ \vert m_{ee}^{(3)} \vert }{ \vert m_{ee}^{(2)} \vert }  < 0,    \hskip1cm
     \lambda_{+}  =   \vert m_{ee}^{(2)} \vert^2 +  \vert m_{ee}^{(3)} \vert^2,   \hskip1cm
     \lambda_{-}  =  0,
  \label{cbbb}
\end{equation}
and $ -{\pi / 4} < \theta < 0$.
This fact shows that, within the approximation up to the second order,
   the isocontour around the point of
   of $ \vert m_{ee} \vert_{\rm min} $ is not an ellipse but a straight line,
\begin{equation}
   0
   +  \frac{  \alpha''^2 }{ \left( \sqrt{ K \over \lambda_+  } \right)^2 } = 1.
   \label{cbbc}
\end{equation}
If one takes into account the higher order terms such as
    $O(  (\beta - { \pi / 2 })^4), O( (\beta - { \pi / 2 })^3 (\alpha + { \pi / 2 }) ),
     O( (\beta - { \pi / 2 })^2 (\alpha + { \pi / 2 })^2)$,
     and $O( (\alpha + { \pi / 2 })^4)$,
     the isocontour would become a closed line. \\
\subsubsection{ The region (B) : 
   $ \vert m_{ee}^{(2)} \vert -   \vert m_{ee}^{(3)} \vert
               <  \vert m_{ee}^{(1)} \vert <  \vert m_{ee}^{(2)} \vert +  \vert m_{ee}^{(3)} \vert $. }
In the region (B), the effective Majorana mass has a minimum value
   $ \vert m_{ee} \vert_{\rm min} $,
\begin{equation}
      \vert m_{ee} \vert_{\rm min} =0.
  \label{cbca}
\end{equation}
The reason why the minimum value becomes zero \cite{rf:Rod2} is explained
   as follows \cite{rf:NunTevFun,rf:Rod3}.
The effective Majorana mass $ \vert m_{ee} \vert $ is given by the absolute value of
   the sum of three complex vectors, $  \vert m_{ee}^{(1)} \vert $,
   $ \vert  m_{ee}^{(2)} \vert e^{2 i \alpha} $, and $  \vert m_{ee}^{(3)} \vert e^{2 i \beta} $.
If these three complex vectors form a triangle, 
   $ \vert m_{ee}^{(1)} \vert +  \vert m_{ee}^{(2)} \vert e^{2 i \alpha}
    +  \vert m_{ee}^{(3)} \vert e^{2 i \beta} = 0$,
   one has $ \vert m_{ee} \vert = 0$ \cite{rf:NunTevFun}.
Now, we look for the necessary condition to form a triangle using the assumption
   $  \vert m_{ee}^{(3)} \vert <  \vert m_{ee}^{(2)} \vert  $ made in section 2.
In the case of $  \vert m_{ee}^{(2)} \vert <  \vert m_{ee}^{(1)} \vert  $,
   the necessary condition is 
   $ \vert m_{ee}^{(1)} \vert <  \vert m_{ee}^{(2)} \vert + \vert m_{ee}^{(3)} \vert $,
   while in the case of $  \vert m_{ee}^{(1)} \vert <  \vert m_{ee}^{(2)} \vert  $,
   the necessary condition is 
   $  \vert m_{ee}^{(2)} \vert <  \vert m_{ee}^{(1)} \vert + \vert m_{ee}^{(3)} \vert $.
Therefore, the necessary condition to form a triangle is
   $  \vert m_{ee}^{(2)} \vert -   \vert m_{ee}^{(3)} \vert
               <  \vert m_{ee}^{(1)} \vert <  \vert m_{ee}^{(2)} \vert +  \vert m_{ee}^{(3)} \vert $,
  and we can show that the effective Majorana mass has a minimum value zero
  in the region (B).
  
Next, we seek the values of Majorana phases $ \beta_c $ and $ \alpha_c $ which
   satisfy $ \vert m_{ee} \vert = 0$ \cite{rf:LinMerRod}.
Since $ \vert m_{ee} \vert $ is invariant under 
   $ (\beta, \alpha) \rightarrow  (-\beta, -\alpha) $,
\begin{equation}
    \vert m_{ee} \vert (\beta, \alpha) =  \vert m_{ee} \vert (-\beta, -\alpha),
  \label{cbcbz}
\end{equation}
we restrict $ \beta_c $ and $ \alpha_c $ within
\begin{equation}
   - {\pi \over 2}   <  \beta_c <  {\pi \over 2},   \hskip1cm
   - {\pi \over 2}   <  \alpha_c  <  0.
  \label{cbcb}
\end{equation}
Solving the equation,
\begin{eqnarray}
   0 & = &    \vert m_{ee}^{(1)} \vert +  \vert m_{ee}^{(2)} \vert e^{2 i \alpha_c}
                        +  \vert m_{ee}^{(3)} \vert e^{2 i \beta_c}                          \nonumber \\
      & = & \left\{  \vert m_{ee}^{(1)} \vert +  \vert m_{ee}^{(2)} \vert \cos 2 \alpha_c
             +  \vert m_{ee}^{(3)} \vert \cos 2 \beta_c \right\}
             + i  \left\{  \vert m_{ee}^{(2)} \vert \sin 2 \alpha_c
             +  \vert m_{ee}^{(3)} \vert \sin 2 \beta_c \right\},
  \label{cbcc}
\end{eqnarray}
we obtain
\begin{equation}
   \cos 2 \alpha_c
   = \frac{  \vert m_{ee}^{(3)} \vert^2 -  \vert m_{ee}^{(1)} \vert^2 -  \vert m_{ee}^{(2)} \vert^2 }
                   { 2 \,  \vert m_{ee}^{(1)} \vert   \vert m_{ee}^{(2)} \vert }, 
  \label{cbcd}
\end{equation}
and then
\begin{equation}
   \cos 2 \beta_c
   = \frac{  \vert m_{ee}^{(2)} \vert^2 -  \vert m_{ee}^{(3)} \vert^2 -  \vert m_{ee}^{(1)} \vert^2 }
                   { 2 \,  \vert m_{ee}^{(1)} \vert   \vert m_{ee}^{(3)} \vert }.
  \label{cbce}
\end{equation}
The Eq.(\ref{cbcd}) and the assumption
   $  \vert m_{ee}^{(3)} \vert <  \vert m_{ee}^{(2)} \vert  $ lead to
   $ - {\pi / 2}   <  \alpha_c  <  - {\pi / 4} $, and then $ \sin 2 \alpha_c < 0$.
From Eq.(\ref{cbcc}), the product of $ \sin 2  \alpha_c $ and $ \sin 2 \beta_c  $
   must be negative, $ \sin 2  \alpha_c  \sin 2 \beta_c < 0 $ \cite{rf:DevKum}.
Consequently, $ \sin 2 \beta_c  $ must be positive and $ 0 <  \beta_c  <  {\pi \over 2} $.
We finally find that $ \beta_c $ and $ \alpha_c $ satisfy Eq.(\ref{cbce}) and Eq.(\ref{cbcd}),
   respectively, and they are within
\begin{equation}
   0   <  \beta_c  <  {\pi \over 2},   \hskip1cm
   - {\pi \over 2}   <  \alpha_c  <  - {\pi \over 4}.
  \label{cbcf}
\end{equation}

We expand $  \vert m_{ee} \vert^2 $ around the point $ (\beta_c, \alpha_c) $ and
   it is approximated by the terms up to second order,
\begin{eqnarray}
   \vert m_{ee} \vert^2
    & \approx &   4  ( \beta - \beta_c )^2  \,  \vert m_{ee}^{(3)} \vert^2
                         + 4  ( \alpha - \alpha_c )^2  \,  \vert m_{ee}^{(2)} \vert^2                   \nonumber  \\
    &  & ~~    + 8  ( \beta - \beta_c )  ( \alpha - \alpha_c ) \biggl\{  \vert m_{ee}^{(2)} \vert   \vert m_{ee}^{(3)} \vert  \cos ( 2 \alpha_c - 2 \beta_c ) \biggr\},
  \label{cbcg}
\end{eqnarray}
or
\begin{equation}
     \left(     \begin{array}{cc}
         \beta'    &   \alpha' 
                 \end{array}
                                 \right)
    \left(     \begin{array}{cc}
        a  &   b         \\
        b  &   c
                 \end{array}
                                 \right)
   \left(     \begin{array}{c}
          \beta'          \\
          \alpha' 
                 \end{array}
                                 \right)
      \approx  K,
     \label{cbch}
\end{equation}
where
\begin{equation}
   \left(     \begin{array}{c}
          \beta'          \\
          \alpha'
                 \end{array}
                                 \right)
   =    \left(  \begin{array}{c}
                     \beta - \beta_c            \\
                     \alpha - \alpha_c
                 \end{array}
                                 \right),
  \label{cbci}
\end{equation}
\begin{eqnarray}
  \hskip3cm
   a   & = &   \vert m_{ee}^{(3)} \vert^2   > 0,                                      \nonumber  \\
   b   & = &  {1 \over 2} \biggl\{  \vert m_{ee}^{(1)} \vert^2 - \vert m_{ee}^{(2)} \vert^2
                          -  \vert m_{ee}^{(3)} \vert^2  \biggr\},                               \nonumber  \\
   c   & = &    \vert m_{ee}^{(2)} \vert^2   > 0,                                   \nonumber  \\
   K  & = &    {1 \over 4}   \vert m_{ee} \vert^2,
  \label{cbcj}
\end{eqnarray}
and then $ a c - b^2 > 0 $ and $ c-a>0 $.
As in section 3.1, Eq.(\ref{cbch}) can be rewritten as
\begin{equation}
     \left(     \begin{array}{cc}
         \beta''    &   \alpha'' 
                  \end{array}
                                 \right)
    \left(     \begin{array}{cc}
        \lambda_{-}  &   0         \\
        0                   &    \lambda_{+}
                 \end{array}
                                 \right)
   \left(     \begin{array}{c}
          \beta''          \\
          \alpha'' 
                 \end{array}
                                 \right)
      \approx  K,
     \label{cbck}
\end{equation}
where
\begin{equation}
    \left(     \begin{array}{c}
          \beta''          \\
          \alpha''
                 \end{array}
                                 \right)
  =   \left(     \begin{array}{cc}
        \cos \theta   &   \sin \theta         \\
        - \sin \theta  &  \cos \theta
                 \end{array}
                                 \right)
     \left(     \begin{array}{c}
          \beta'           \\
          \alpha' 
                 \end{array}
                                 \right)
     =
     \left(     \begin{array}{cc}
        \cos \theta   &   \sin \theta         \\
        - \sin \theta  &  \cos \theta
                 \end{array}
                                 \right)
     \left(     \begin{array}{c}
         ( \beta - \beta_c )            \\
         ( \alpha - \alpha_c )
                 \end{array}
                                 \right),
  \label{cbcl}
\end{equation}
and
\begin{eqnarray}
   \tan \theta & =& \frac{ 1 }{  \vert m_{ee}^{(1)} \vert^2 - \vert m_{ee}^{(2)} \vert^2
                          -  \vert m_{ee}^{(3)} \vert^2 }                
         \biggl[  \vert m_{ee}^{(2)} \vert^2  - \vert m_{ee}^{(3)} \vert^2   \biggr.      \nonumber  \\
     & &  \biggl.  - \sqrt{  \vert m_{ee}^{(1)} \vert^4 + 2  \vert m_{ee}^{(2)} \vert^4
               + 2  \vert m_{ee}^{(3)} \vert^4 - 2  \vert m_{ee}^{(1)} \vert^2
                  (  \vert m_{ee}^{(2)} \vert^2 +  \vert m_{ee}^{(3)} \vert^2 ) } \, \biggr],         \nonumber  \\
    \tan 2 \theta & =&   \frac{  \vert m_{ee}^{(1)} \vert^2 - \vert m_{ee}^{(2)} \vert^2
                     -  \vert m_{ee}^{(3)} \vert^2 }{  \vert m_{ee}^{(3)} \vert^2 -  \vert m_{ee}^{(2)} \vert^2  },
                                                                                           \nonumber  \\
  \lambda_{\pm} &=&  {1 \over 2} \biggl[   \vert m_{ee}^{(2)} \vert^2  + \vert m_{ee}^{(3)} \vert^2   \biggr.          \nonumber  \\
     &    &   \biggl.   \pm \sqrt{   \vert m_{ee}^{(1)} \vert^4 + 2  \vert m_{ee}^{(2)} \vert^4
               + 2  \vert m_{ee}^{(3)} \vert^4 - 2  \vert m_{ee}^{(1)} \vert^2
                  (  \vert m_{ee}^{(2)} \vert^2 +  \vert m_{ee}^{(3)} \vert^2 )          } \,  \biggr].
  \label{cbcm}
\end{eqnarray}
The sign of the angle $\theta$ depends on the magnitude of $ \vert m_{ee}^{(1)} \vert $.
When the $ \vert m_{ee}^{(1)} \vert $ satisfies
   $ \sqrt{  \vert m_{ee}^{(2)} \vert^2 +  \vert m_{ee}^{(3)} \vert^2 } 
       <  \vert m_{ee}^{(1)} \vert <  \vert m_{ee}^{(2)} \vert +  \vert m_{ee}^{(3)} \vert $,
   the parameter $b$ becomes positive and we have $ - {\pi / 4} < \theta < 0 $.
On the other hand, when the $ \vert m_{ee}^{(1)} \vert $ satisfies
   $  \vert m_{ee}^{(2)} \vert -  \vert m_{ee}^{(3)} \vert   <  \vert m_{ee}^{(1)} \vert
          <  \sqrt{  \vert m_{ee}^{(2)} \vert^2 +  \vert m_{ee}^{(3)} \vert^2 }  $,
   the parameter $b$ becomes negative and we have $ 0 < \theta <  {\pi / 4} $.
In the region (B), 
   $ \vert m_{ee}^{(2)} \vert -   \vert m_{ee}^{(3)} \vert
               <  \vert m_{ee}^{(1)} \vert <  \vert m_{ee}^{(2)} \vert +  \vert m_{ee}^{(3)} \vert $,
   the isocontour around the point of the minimum $ \vert m_{ee} \vert_{\rm min} =0$
   becomes the ellipse represented by the following equation,
\begin{equation}
   \frac{  \beta''^2 }{ \left( \sqrt{ K \over \lambda_-  } \right)^2 }
   +  \frac{  \alpha''^2 }{ \left( \sqrt{ K \over \lambda_+  } \right)^2 } = 1,
   \label{cbcn}
\end{equation}
where the third order of two variables $ (\beta - \beta_c) $ and $ (\alpha - \alpha_c) $
   or higher orders have been neglected.
The center of this ellipse is the point $ (\beta_c, \alpha_c) $.
From the invariance of $ \vert m_{ee} \vert $, Eq.(\ref{ab}) and Eq.(\ref{cbcbz}),
   the ellipse with the center $ (\pm \beta_c + n \pi, \,  \pm \alpha_c  + m \pi)$
   is distributed in the $ \beta \alpha-$plane.
   
The expressions of $ \cos 2 \alpha_c, \, \cos 2 \beta_c, \, \tan \theta $,
   and $\lambda_{\pm}$ are very complicated.
However, when the value of the lightest neutrino mass $m_1$
   ($  \vert m_{ee}^{(1)} \vert = m_1  \vert U_{e 1} \vert^2 $) satisfies
   $ \vert m_{ee}^{(1)} \vert^2 = \vert m_{ee}^{(2)} \vert^2 + \vert m_{ee}^{(3)} \vert^2 $,
   three complex vectors, $ \vert m_{ee}^{(1)} \vert  $,
   $ \vert m_{ee}^{(2)} \vert e^{ 2 i \beta}$, and $ \vert m_{ee}^{(3)} \vert  e^{ 2 i \alpha }$
   form a right-angle triangle, and the expressions become very simple,
\begin{equation}
  \theta = 0, \hskip0.5cm  \lambda_{-} = a =   \vert m_{ee}^{(3)} \vert^2,  \hskip0.5cm
               \lambda_{+} =  c = \vert m_{ee}^{(2)} \vert^2,  \hskip0.5cm  
               \sqrt{ \lambda_{-} \over \lambda_{+} }
                  =  \frac{  \vert m_{ee}^{(3)} \vert }{  \vert m_{ee}^{(2)} \vert } < 1,
  \label{cbco}
\end{equation}
and
\begin{equation}
  \cos (2 \alpha_c )  =   -   \frac{  \vert m_{ee}^{(2)} \vert }{  \vert m_{ee}^{(1)} \vert },
  \hskip1cm 
    \cos (2 \beta_c )  =   -   \frac{  \vert m_{ee}^{(3)} \vert }{  \vert m_{ee}^{(1)} \vert }.
    \label{cbcp}
\end{equation}
\subsubsection{ On the boundary between the region (B) and region (C). }
We consider the boundary between the region (B) and region (C),
\begin{equation}
   \vert m_{ee}^{(1)} \vert =  \vert m_{ee}^{(2)} \vert  -  \vert m_{ee}^{(3)} \vert.
  \label{cbda}
\end{equation}
On this boundary, the effective Majorana mass has a minimum value
   $ \vert m_{ee} \vert_{\rm min} =0 $ at
   $ (\beta, \alpha) =  (0,  - { \pi / 2 }) $.
When the $  \vert m_{ee} \vert^2  $ is expanded around the point 
   $ (\beta, \alpha) = (0,  - { \pi / 2 })  $, there appears no odd order term
   on the two variables $ \beta $ and $ (\alpha + \pi / 2)$ in the expansion.
The expansion up to the second order takes the same form to Eq.(\ref{cbcg}).
Because of $c-a > 0$ and $ ac-b^2 = 0$ on this boundary, one has
\begin{equation}
     \tan \theta  =   \frac{ \vert m_{ee}^{(3)} \vert }{ \vert m_{ee}^{(2)} \vert }  < 1,   \hskip0.5cm
     \lambda_{+}  =   \vert m_{ee}^{(2)} \vert^2 +  \vert m_{ee}^{(3)} \vert^2,   \hskip0.5cm
     \lambda_{-}  =  0,
  \label{cbdb}
\end{equation}
and $0 < \theta <  {\pi / 4}$.
This fact shows that, within the approximation up to the second order,
   the isocontour around the point of
   of $ \vert m_{ee} \vert_{\rm min} $ is not an ellipse but a straight line,
\begin{equation}
   0
   +  \frac{  \alpha''^2 }{ \left( \sqrt{ K \over \lambda_+  } \right)^2 } = 1.
   \label{cbdc}
\end{equation}
If one takes into account the higher order terms such as
   $O(\beta^4), O(\beta^3 ( \alpha + { \pi / 2 }) ),
      O(\beta^2 ( \alpha + { \pi / 2 })^2)$,
      and $O( (\alpha + { \pi / 2 })^4)$,
     the isocontour would become a closed line.
\subsubsection{ The region (C) : 
    $    \vert m_{ee}^{(1)} \vert < \vert m_{ee}^{(2)} \vert - \vert m_{ee}^{(3)} \vert  $}
In the region (C), the effective Majorana mass has a minimum value
   $ \vert m_{ee} \vert_{\rm min} $,
\begin{equation}
    \vert m_{ee} \vert_{\rm min}
     = \vert  \vert m_{ee}^{(1)} \vert -  \vert m_{ee}^{(2)} \vert
          +  \vert m_{ee}^{(3)} \vert  \vert,
  \label{cbea}
\end{equation}
at the point $ (\beta, \alpha) = (0,  - { \pi / 2 }) $.
We expand $  \vert m_{ee} \vert^2  $ around this point and it is approximated by the terms
   up to second order,
\begin{eqnarray}
   \vert m_{ee} \vert^2
    & \approx &  \vert m_{ee} \vert_{\rm min}^2
                   + 4  \beta^2 \biggl\{  - (   \vert m_{ee}^{(1)} \vert - \vert m_{ee}^{(2)} \vert )
                                     \vert m_{ee}^{(3)} \vert \biggr\}                             \nonumber  \\
    &  & ~~  + 4 ( \alpha + { \pi \over 2 })^2 \biggl\{ (   \vert m_{ee}^{(1)} \vert + \vert m_{ee}^{(3)} \vert )
                                     \vert m_{ee}^{(2)} \vert \biggr\}                               \nonumber  \\
    &  & ~~    - 8  \beta  ( \alpha + { \pi \over 2 })  \vert m_{ee}^{(2)} \vert   \vert m_{ee}^{(3)} \vert,
  \label{cbeb}
\end{eqnarray}
or
\begin{equation}
     \left(   \begin{array}{cc}
                  \beta'   &    \alpha'
                 \end{array}
                                 \right)
    \left(     \begin{array}{cc}
        a  &   b         \\
        b  &   c
                 \end{array}
                                 \right)
   \left(     \begin{array}{c}
                  \beta'           \\
                  \alpha'
                 \end{array}
                                 \right)
      \approx  K,
     \label{cbec}
\end{equation}
where
\begin{equation}
   \left(   \begin{array}{c}
                  \beta'          \\
                  \alpha' 
              \end{array}
                                 \right)
     =   \left(     \begin{array}{c}
           \beta             \\
          \alpha + { \pi \over 2 }
                 \end{array}
                                 \right),
  \label{cbed}
\end{equation}
\begin{eqnarray}
   \hskip2.5cm
   a   & = &   - (   \vert m_{ee}^{(1)} \vert - \vert m_{ee}^{(2)} \vert )
                                 \vert m_{ee}^{(3)} \vert   > 0,                                      \nonumber  \\
   b   & = &   -  \vert m_{ee}^{(2)} \vert   \vert m_{ee}^{(3)} \vert  \hskip1.3cm   < 0,      \nonumber  \\
   c   & = &   (   \vert m_{ee}^{(1)} \vert + \vert m_{ee}^{(3)} \vert )
                                 \vert m_{ee}^{(2)} \vert   > 0,                                   \nonumber  \\
   K  & = &    {1 \over 4}  \biggl\{   \vert m_{ee} \vert^2 - \vert m_{ee} \vert_{\rm min}^2
                                                                                                                   \biggr\} > 0,
  \label{cbee}
\end{eqnarray}
and then $ a c - b^2 > 0 $ and $ c-a>0 $.
As in section 3.1, Eq.(\ref{cbec}) can be rewritten as
\begin{equation}
     \left(  \begin{array}{cc}
                \beta''   &    \alpha''
              \end{array}
                                 \right)
    \left(     \begin{array}{cc}
        \lambda_{-}  &   0         \\
        0                   &    \lambda_{+}
                 \end{array}
                                 \right)
   \left(   \begin{array}{c}
                \beta''          \\
                \alpha''
             \end{array}
                                 \right)
      \approx  K,
     \label{cbef}
\end{equation}
where
\begin{equation}
    \left(   \begin{array}{c}
                \beta''          \\
                \alpha''
             \end{array}
                                 \right)
  =  \left(     \begin{array}{cc}
        \cos \theta   &   \sin \theta         \\
        - \sin \theta  &  \cos \theta
                 \end{array}
                                 \right)
     \left(   \begin{array}{c}
                   \beta'           \\
                   \alpha'
               \end{array}
                                 \right)
  =
     \left(     \begin{array}{cc}
        \cos \theta   &   \sin \theta         \\
        - \sin \theta  &  \cos \theta
                 \end{array}
                                 \right)
     \left(     \begin{array}{c}
             \beta            \\
            \alpha + { \pi \over 2 }
                 \end{array}
                                 \right),
  \label{cbeg}
\end{equation}
and
\begin{eqnarray}
   \tan \theta & =&  \frac{ - 1 }{ 2    \vert m_{ee}^{(2)} \vert   \vert m_{ee}^{(3)} \vert }
                         \biggl[    \vert m_{ee}^{(1)} \vert  (  \vert m_{ee}^{(2)} \vert  + \vert m_{ee}^{(3)} \vert  )    \biggr.                \nonumber  \\
     &    &  ~~   \left.   -  \sqrt{  \vert m_{ee}^{(1)} \vert^2 ( \vert m_{ee}^{(3)} \vert +  \vert m_{ee}^{(2)} \vert )^2
               + 4 (  \vert m_{ee}^{(2)} \vert \vert m_{ee}^{(3)} \vert )^2  } \,  \right] > 0,   \nonumber    \\ 
    \tan 2 \theta & =&   \frac{2    \vert m_{ee}^{(2)} \vert   \vert m_{ee}^{(3)} \vert }
                          {   \vert m_{ee}^{(1)} \vert (   \vert m_{ee}^{(2)} \vert +  \vert m_{ee}^{(3)} \vert ) }>0,
                                                              \nonumber  \\
  \lambda_{\pm} &=&  {1 \over 2} \left[  \biggl\{
                   -  \vert m_{ee}^{(1)} \vert  \vert m_{ee}^{(3)} \vert  + 2  \vert m_{ee}^{(2)} \vert   \vert m_{ee}^{(3)} \vert
                     +   \vert m_{ee}^{(1)} \vert  \vert m_{ee}^{(2)} \vert  \biggr\}    \right.                \nonumber  \\
     &    &  ~~   \left.   \pm \sqrt{  \vert m_{ee}^{(1)} \vert^2 ( \vert m_{ee}^{(3)} \vert +  \vert m_{ee}^{(2)} \vert )^2
               + 4 (  \vert m_{ee}^{(2)} \vert \vert m_{ee}^{(3)} \vert )^2  } \,  \right].
  \label{cbeh}
\end{eqnarray}
Because of $ \tan \theta > 0 $ and $  \tan 2 \theta > 0 $, one has
   $ 0 < \theta <  {\pi / 4} $.
In the region (C), 
   $ \vert m_{ee}^{(1)} \vert < \vert m_{ee}^{(2)} \vert - \vert m_{ee}^{(3)} \vert $,
   the isocontour around the point of the minimum,
   $ \vert m_{ee} \vert_{\rm min}
     = \vert m_{ee}^{(2)} \vert -  \vert m_{ee}^{(3)} \vert
          -  \vert m_{ee}^{(1)} \vert $
  becomes the ellipse represented by the following equation,
\begin{equation}
   \frac{ \beta''^2 }{ \left( \sqrt{ K \over \lambda_-  } \right)^2 }
   +  \frac{ \alpha''^2 }{ \left( \sqrt{ K \over \lambda_+  } \right)^2 } = 1,
   \label{cbei}
\end{equation}
where the fourth order of two variables $ \beta $ and $ (\alpha + \pi / 2)$ or higher
   orders have been neglected.
This ellipse has a center at $ (\beta, \alpha) =  (0,  - { \pi / 2 }) $.
The direction of the $\beta''$ axis is produced by a counterclockwise rotation of the
   $\beta$ axis by the angle $ \theta (>0)$.
   
Although the expressions of $\lambda_{\pm}$ or $ \tan \theta $ are very complicated,
   these expressions become simple when the value of the lightest neutrino mass
   $m_1$ ($  \vert m_{ee}^{(1)} \vert = m_1  \vert U_{e 1} \vert^2 $) satisfies,
\begin{equation}
   \vert m_{ee}^{(3)} \vert  >>  \vert m_{ee}^{(1)} \vert.
  \label{cbej}
\end{equation}
Owing to this inequality, we can neglect the terms of order
   $O(\vert m_{ee}^{(1)} \vert^2)$, then the angle $ \theta $ and the ratio of the minor
   axis to the major axis $ \sqrt{  \lambda_{-} /  \lambda_{+} } $ become
\begin{eqnarray}
   \theta & \approx & { \pi \over 4 } - \frac{  ( \vert m_{ee}^{(2)} \vert
                                  + \vert m_{ee}^{(3)} \vert ) }
                                  { 4 \,  \vert m_{ee}^{(2)} \vert  \vert m_{ee}^{(3)} \vert }
                                    \vert m_{ee}^{(1)} \vert,               \nonumber \\
   \sqrt{ \lambda_{-} \over \lambda_{+} }
              & \approx &  \sqrt{   \frac{  ( \vert m_{ee}^{(2)} \vert -  \vert m_{ee}^{(3)} \vert ) }
                                   { 4  \vert m_{ee}^{(2)} \vert   \vert m_{ee}^{(3)} \vert }
                                     \vert m_{ee}^{(1)} \vert  }.
  \label{cbek}
\end{eqnarray}
In the region where the condition Eq.(\ref{cbej}) is satisfied,
   the angle $\theta$ increases and the ratio
   $ \sqrt{  \lambda_{-} /  \lambda_{+} } $ decreases
   when the value of the lightest neutrino mass $m_1$ is decreased.
When the size of the major axis of the ellipse is $ \pi / 4$, the effective Majorana 
   mass becomes
\begin{eqnarray}
      \vert m_{ee} \vert & \approx &  \vert m_{ee} \vert_{\rm min} 
                                           + {\pi^2 \over 64}   \vert m_{ee}^{(1)} \vert.
      \label{cbekx}                                       
\end{eqnarray}
In the $ m_1 \rightarrow 0 $ limit, we have
\begin{equation}
   \theta  \rightarrow   { \pi \over 4} ,  \hskip1cm
   \sqrt{ \lambda_{-} \over \lambda_{+} }   \rightarrow   0,   \hskip1cm (m_1 \rightarrow 0).
  \label{cbel}
\end{equation}
These behavior in the $ m_1 \rightarrow 0 $ limit resembles those of Eq.(\ref{caw})
   in the case of the ellipse representing the isocontour around the point of the
   maximum $  \vert m_{ee} \vert  $.
\subsection{ Region except extremal values of $ \vert m_{ee} \vert  $.}
In the preceding subsections, we have obtained the isocontour around the point of
   the maximum or minimum $ \vert m_{ee} \vert $ in the $ \beta \alpha-$plane.
What are the isocontours not around the point of the maximum or minimum
   $ \vert m_{ee} \vert  $?
We shall obtain such isocontours in this subsection.
In the normal mass ordering case, the $ \vert m_{ee}^{(1)} \vert $ changes from zero to
   the maximum of $  \vert m_{ee}^{(i)} \vert \, (i=1,2,3) $ when the value of 
  the lightest neutrino mass $m_1$ is varied.
If the values of $m_1$
   (or $  \vert m_{ee}^{(1)} \vert = m_1  \vert U_{e 1} \vert^2 $) is in some regions,
   one can obtain the approximated equations analytically which represent the
   isocontours not around the point of $ \vert m_{ee} \vert_{\rm max}$ or
   $ \vert m_{ee} \vert_{\rm min}$.
In the following, we give such three regions of $ \vert m_{ee}^{(1)} \vert $ and derive
   the approximated equation representing the isocontour in each three cases.
\subsubsection{ The first case: 
      $  \vert m_{ee}^{(1)} \vert >   \vert m_{ee}^{(2)} \vert >>  \vert m_{ee}^{(3)} \vert. $}
The first case is that the value of the lightest neutrino mass $m_1$
   ($ = \vert m_{ee}^{(1)} \vert /  \vert U_{e 1} \vert^2 $) is large enough to satisfy
    the following relation,
\begin{equation}
      \vert m_{ee}^{(1)} \vert >   \vert m_{ee}^{(2)} \vert >>  \vert m_{ee}^{(3)} \vert.
  \label{ccaa}
\end{equation}
In order to satisfy this relation, $m_1^2$ should be much larger than
   $  \bigtriangleup m_{\odot}^2 $,
   $ m_1^2  >>   \bigtriangleup m_{\odot}^2 $.
In this first case, $ \vert m_{ee}^{(3)} \vert $ is much smaller than 
   $ \vert m_{ee}^{(1)} \vert $ and $ \vert m_{ee}^{(2)} \vert $.
At first, let $ \vert m_{ee}^{(3)} \vert $ equal zero and the Majorana phase $ \alpha_0 $
   satisfies,
   $  \vert m_{ee} \vert = \left\vert \, \vert m_{ee}^{(1)} \vert
         +  \vert m_{ee}^{(2)} \vert e^{2 i \alpha_0} \right\vert $, or
\begin{equation}
 \cos ( 2 \alpha_0 ) = \frac{  \vert m_{ee} \vert^2 -  \vert m_{ee}^{(1)} \vert^2 - \vert m_{ee}^{(2)} \vert^2 }
                                  { 2 \vert m_{ee}^{(1)} \vert  \vert m_{ee}^{(2)} \vert }.
  \label{ccab}
\end{equation}
The other phase $ \beta $ can not be determined at all.
In practice the $ \vert m_{ee}^{(3)} \vert$ is not zero, and we will obtain the isocontour
   of $ \vert m_{ee} \vert $ by the method of perturbation with $ \vert m_{ee}^{(3)} \vert$,
   or more formally with
   $  \vert m_{ee}^{(3)} \vert  /   \sqrt{  \vert m_{ee}^{(1)} \vert  \vert m_{ee}^{(2)} \vert }  $.
We rewrite
\begin{eqnarray}
   \vert m_{ee} \vert
 &=&  \vert \, \vert m_{ee}^{(1)} \vert +  \vert m_{ee}^{(2)} \vert e^{2 i \alpha}
   +  \vert m_{ee}^{(3)} \vert e^{2 i \beta} \, \vert   \nonumber \\
 & \equiv & \sqrt{ \vert m_{ee}^{(1)} \vert  \vert m_{ee}^{(2)} \vert } ~
       \vert X_1+X_2 \, e^{2 i \alpha}+X_3 \, e^{2 i \beta} \vert,
  \label{ccac}
\end{eqnarray}
or
\begin{equation}
   \frac{ \vert m_{ee} \vert^2 }{ \vert m_{ee}^{(1)} \vert  \vert m_{ee}^{(2)} \vert }
     =  \vert X_1+X_2 \, e^{2 i \alpha}+X_3 \, e^{2 i \beta} \vert^2,
  \label{ccad}
\end{equation}
where
\begin{equation}
   X_1 \equiv \sqrt{ \vert m_{ee}^{(1)} \vert \over  \vert m_{ee}^{(2)} \vert }, \hskip1cm
   X_2 \equiv \sqrt{ \vert m_{ee}^{(2)} \vert \over  \vert m_{ee}^{(1)} \vert },\hskip1cm
   X_3 \equiv   \frac{ \vert m_{ee}^{(3)} \vert  } { \sqrt{ \vert m_{ee}^{(1)} \vert  \vert m_{ee}^{(2)} \vert } } ,
     \label{ccae}
\end{equation}
and $ X_1 > X_2 >> X_3 $, $ X_1 X_2 =1$.

If $ X_3 (<<1)$ in the right-handed side of Eq.(\ref{ccad}) is neglected, the phase
   $ \alpha = \alpha_0 $ satisfies
\begin{equation}
   \frac{ \vert m_{ee} \vert^2 }{ \vert m_{ee}^{(1)} \vert  \vert m_{ee}^{(2)} \vert }
     =  \vert X_1+X_2 \, e^{2 i \alpha_0}  \vert^2,
   \label{ccaf}
\end{equation}
or
\begin{equation}
 \cos ( 2 \alpha_0 )
    = {1 \over 2} \left[
    \frac{  \vert m_{ee} \vert^2  }{ \vert m_{ee}^{(1)} \vert  \vert m_{ee}^{(2)} \vert }
     - X_1^2 - X_2^2 \right].
  \label{ccag}
\end{equation}
Now, we seek the $\alpha$ by the perturbation with $X_3$, and we expand $ \alpha $ as
\begin{equation}
 \alpha  =  \sum_{n=0}^{\infty} \alpha_n X_3^n    
            =   \alpha_0 + \alpha_1 X_3 + \alpha_2 X_3^2+ \cdots.
  \label{ccah}
\end{equation}
Then, it leads
\begin{eqnarray}
  & &   \vert X_1+X_2 \, e^{2 i \alpha}+X_3 \, e^{2 i \beta} \vert^2  \nonumber  \\
  & = & \vert  X_1 + X_2 \, e^{2 i ( \alpha_0 + \alpha_1 X_3 + \alpha_2 X_3^2+ \cdots ) }
                + X_3 \, e^{ 2 i \beta} \vert^2  \nonumber  \\
  & = &  \vert  a_0 + a_1 \, X_3 + a_2 \, X_3^2 + \cdots  \vert^2,
  \label{ccai}
\end{eqnarray}
where
\begin{eqnarray}
  a_0  &\equiv&  X_1+X_2 \, e^{2 i \alpha_0},  \nonumber  \\
  a_1  &\equiv&  2 i \alpha_1 X_2  \, e^{2 i \alpha_0} +  \, e^{2 i \beta},  \nonumber  \\
  a_2  &\equiv&  2 X_2  \,( i  \alpha_2 -  \alpha_1^2 ) e^{2 i \alpha_0},
     \label{ccaj}
\end{eqnarray}
and so on.
Owing to the definition of $ \alpha_0$, the $a_0$ satisfies
\begin{equation}
  \frac{ \vert m_{ee} \vert^2 }
    {  \vert m_{ee}^{(1)} \vert  \vert m_{ee}^{(2)} \vert }
     =  \vert  a_0 \vert^2.
  \label{ccak}
\end{equation}
When the order of $ X_3^2 $ is neglected in Eq(\ref{ccai}), one has from Eq.(\ref{ccad}),
\begin{equation}
   \frac{ \vert m_{ee} \vert^2 }{ \vert m_{ee}^{(1)} \vert  \vert m_{ee}^{(2)} \vert }
     =  \vert  a_0 + a_1 \, X_3 + O(X_3^2)  \vert^2
     =  \vert a_0 \vert^2 + 2 {\rm Re} ( a_0^{\ast} a_1 ) X_3 + O(X_3^2).
  \label{ccal}
\end{equation}
The $ \alpha_1$ can be determined by
\begin{equation}
  {\rm Re} ( a_0^{\ast} a_1 ) = 0,
  \label{ccam}
\end{equation}
and the result is
\begin{eqnarray}
  \alpha_1 
    & = &   { 1 \over 2 \sin ( 2 \alpha_0 ) }
              \left\{ ( {\rm Re} \, a_0 ) \cos ( 2 \beta ) +  ( {\rm Im} \, a_0 ) \sin ( 2 \beta ) \right\}
                                                                                \nonumber \\
     & = &   { 1 \over 2 \sin ( 2 \alpha_0 ) } 
                     \left\{ \vert a_0 \vert  \,  \cos ( 2 \beta + \phi ) \right\}       \nonumber \\
     & = &    { 1 \over 2 \sin ( 2 \alpha_0 ) }
                    \sqrt { X_1^2 + X_2^2 + 2 \cos (  2 \alpha_0 ) } \,  \cos ( 2 \beta + \phi )
                                                                                                   \nonumber \\
     & = &  { 1 \over 2 \sin ( 2 \alpha_0 ) }
                 \frac{ \vert m_{ee} \vert }
                     { \sqrt{  \vert m_{ee}^{(1)} \vert  \vert m_{ee}^{(2)} \vert } }
                     \,  \cos ( 2 \beta + \phi ),                 
  \label{ccan}
\end{eqnarray}
where
\begin{equation}
  \tan \phi = - \frac{  {\rm Im} \, a_0 }{  {\rm Re} \, a_0 }
                = - \frac{ X_2 \sin ( 2 \alpha_0 ) }{X_1 +  X_2 \cos ( 2 \alpha_0 ) }.
  \label{ccao}
\end{equation}
In the same manner, the coefficient, $ \alpha_2 $ of the second order of $ X_3^2 $
  can be obtained,
\begin{equation}
  \alpha_2 = { 1 \over 4 \sin ( 2 \alpha_0 ) }
                 \biggl\{   1 - 4 \alpha_1^2 \cos ( 2 \alpha_0 ) + 4 \alpha_1 X_2
                       \sin ( 2 \beta - 2 \alpha_0 ) \biggr\}.
  \label{ccap}
\end{equation}
In this expansion, it should be $ \sin 2 \alpha_0 \ne 0$.
This shows that we can not use this expansion near
   $ \alpha_0 = 0$ or $ \alpha_0 =  - \pi /2 $.
When $ \alpha_0 \approx 0$, it becomes $ \cos ( 2 \alpha_0 ) \approx 1$, and
   $  \vert m_{ee} \vert \approx  \vert m_{ee}^{(1)} \vert +  \vert m_{ee}^{(2)} \vert 
        \approx  \vert m_{ee} \vert_{\rm max} $.
When $ \alpha_0 \approx - \pi /2$, it becomes $ \cos ( 2 \alpha_0 ) \approx -1$, and
   $  \vert m_{ee} \vert \approx  \sqrt{ \vert m_{ee}^{(1)} \vert^2 
        +  \vert m_{ee}^{(2)} \vert^2 - 2  \vert m_{ee}^{(1)} \vert   \vert m_{ee}^{(2)} \vert }
       =  \vert m_{ee}^{(1)} \vert - \vert m_{ee}^{(2)} \vert \approx  
        \vert m_{ee} \vert_{\rm min} $.
From these, we understand that the approximation using 
  the expansion in $X_3$ does not work near the
   maximum value $  \vert m_{ee} \vert_{\rm max} $ or the minimum value
   $   \vert m_{ee} \vert_{\rm min} $.

Since $ X_3 $ is very small in this case, we only take the first order $ O(X_3) $,
   and the relation between $\alpha$ and $\beta$ becomes
\begin{eqnarray}
  \alpha  &=& \alpha_0  +  X_3 \,
                  { 1 \over 2 \sin ( 2 \alpha_0 ) }
                     \sqrt { X_1^2 + X_2^2 + 2 \cos (  2 \alpha_0 ) } \,  \cos ( 2 \beta + \phi )
                      +  O(X_3^2)                                                                  \nonumber \\
             &=& \alpha_0 +  X_3 \,
                      { 1 \over 2 \sin ( 2 \alpha_0 ) }
                      \frac{ \vert m_{ee} \vert }
                      { \sqrt{  \vert m_{ee}^{(1)} \vert  \vert m_{ee}^{(2)} \vert } }
                        \,  \cos ( 2 \beta + \phi )      
                                + O(X_3^2)                               \nonumber  \\
            &=& \alpha_0 +  \left( {m_3 \over m_1 m_2} \right)
                    \frac{  \vert m_{ee} \vert }{ 2 \sin ( 2 \alpha_0 ) }    
                    \frac{ \vert  U_{e 3} \vert^2 }{ \vert U_{e 1} \vert^2   \vert U_{e 2} \vert^2 }   
                     \cos ( 2 \beta + \phi )  + O(X_3^2).
    \label{ccaq}
\end{eqnarray}
The $ \alpha $ is the sum of the constant $ \alpha_0 $ and the cosine
   function of $\beta$ with the period $\pi$ which function is multiplied
   by the small quantity.
   
For instance, consider the case that $ \alpha_0 = - \pi / 4 $.
From Eq.(\ref{ccab}), the effective Majorana mass takes the value
   $  \vert m_{ee} \vert = \sqrt{  \vert m_{ee}^{(1)} \vert^2 +  \vert m_{ee}^{(2)} \vert^2 } $
   and the equation of the isocontour of this $ \vert m_{ee} \vert $ becomes
\begin{equation}
  \alpha = - { \pi \over 4 } 
                    - \frac{ \sqrt{  \vert m_{ee}^{(1)} \vert^2 +  \vert m_{ee}^{(2)} \vert^2 }  }
                      { 2 \vert m_{ee}^{(1)} \vert  \vert m_{ee}^{(2)} \vert }
                       \vert m_{ee}^{(3)} \vert  \,  \cos ( 2 \beta + \phi )      
                                + O(X_3^2),
   \label{ccar}
\end{equation}
where $  \tan \phi = \vert m_{ee}^{(2)} \vert / \vert m_{ee}^{(1)} \vert  < 1$.
When the mass $m_1$ decreases, the factor $ m_3 / (m_1 m_2) $ in the coefficient
   of $ \cos (2 \beta + \phi) $ increases.
Thereby, the contribution of the second term in the right-handed side of Eq.(\ref{ccar})
   becomes larger when the value of the lightest neutrino mass $m_1$ decreases.
While the constraint Eq.(\ref{ccaq}) can not determine the value of $\beta$, it restricts
   the value of $\alpha$ within a rather narrow range.
To see it quantitatively, the second term in the right-handed side of Eq.(\ref{ccaq})
   is estimated as
\begin{equation}
   \left\vert  { 1 \over 2 \sin ( 2 \alpha_0 ) }
                \frac{  \vert m_{ee}^{(3)} \vert }
                      {  \vert m_{ee}^{(1)} \vert  \vert m_{ee}^{(2)} \vert  }
                      \vert m_{ee} \vert   \,  \cos ( 2 \beta + \phi )      \right\vert
    \le  { 1 \over 2 \vert  \sin ( 2 \alpha_0 ) \vert }
                \frac{  \vert m_{ee}^{(3)} \vert }
                      {  \vert m_{ee}^{(1)} \vert  \vert m_{ee}^{(2)} \vert  }
                      \vert m_{ee} \vert.
  \label{ccas}
\end{equation}
For instance, again consider the case of $ \alpha_0 = - \pi / 4 $ and we have
\begin{equation}
   \left\vert  { 1 \over 2 \sin ( 2 \alpha_0 ) }
                \frac{  \vert m_{ee}^{(3)} \vert }
                      {  \vert m_{ee}^{(1)} \vert  \vert m_{ee}^{(2)} \vert  }
                      \vert m_{ee} \vert   \,  \cos ( 2 \beta + \phi )      \right\vert
    \le  { 1 \over 2 }
                \frac{  \sqrt{  \vert m_{ee}^{(1)} \vert^2 +  \vert m_{ee}^{(2)} \vert^2 }  }
                        {  \vert m_{ee}^{(1)} \vert  \vert m_{ee}^{(2)} \vert  }   
              \vert m_{ee}^{(3)} \vert,
  \label{ccat}
\end{equation}
which is small because of
   $  \vert m_{ee}^{(1)} \vert >   \vert m_{ee}^{(2)} \vert >>  \vert m_{ee}^{(3)} \vert $.
When the value of the lightest neutrino mass $m_1$ decreases, the range restricting
   the value of $\alpha$ enlarges.
\subsubsection{ The second case;
        $  \vert m_{ee}^{(1)} \vert  >  \vert m_{ee}^{(2)} \vert  >  \vert m_{ee}^{(3)} \vert   $.}
The second case is that the lightest neutrino mass $m_1$ has the value satisfying
\begin{equation}
   \vert m_{ee}^{(1)} \vert  > \vert m_{ee}^{(2)} \vert  >  \vert m_{ee}^{(3)} \vert.
  \label{ccba}
\end{equation}
This value of $m_1$ is not so large as that in the first case
   (the proceeding sub-subsection). 
In order to satisfy the inequality, $  \vert m_{ee}^{(1)} \vert  >  \vert m_{ee}^{(2)} \vert $,
   the $m_1$ should be
\begin{equation}
   m_1^2  >  \frac{  \bigtriangleup m_{\odot}^2 }
                         {  \frac{  \vert U_{e 1} \vert^4 }{  \vert U_{e 2} \vert^4 } - 1 }.
  \label{ccbb}
\end{equation}
If one puts $ \sin^2 \theta_{12} = 0.30 $ for the moment, it becomes
\begin{equation}
       \frac{ 1 }{  \frac{  \vert U_{e 1} \vert^4 }{  \vert U_{e 2} \vert^4 } - 1 }
       \approx 0.23.
\end{equation}
Although $ \vert m_{ee}^{(3)} \vert $ is smaller than $  \vert m_{ee}^{(1)} \vert  $ and
   $ \vert m_{ee}^{(2)} \vert $, the $ X_3 $ in this second case is not so small as that
   in the first case.
When we take into account the second order of $X_3$ in Eq.(\ref{ccaq}), the $\alpha$
   is represented as 
\begin{equation}
  \alpha = \alpha_0 +  X_3 { 1 \over 2 \sin ( 2 \alpha_0 ) }
                   \sqrt { X_1^2 + X_2^2 + 2 \cos (  2 \alpha_0 ) } \,  \cos ( 2 \beta + \phi )
                  +  \alpha_2  X_3^2 + O(X_3^3),
   \label{ccbc}
\end{equation}
where the coefficient, $ \alpha_2 $, is from Eq.(\ref{ccap}),
\begin{eqnarray}
  \alpha_2 
  & = & { 1 \over 4 \sin ( 2 \alpha_0 ) }
             \biggl\{   1 - 4 \alpha_1^2 \cos ( 2 \alpha_0 ) + 4 \alpha_1 X_2
              \sin ( 2 \beta - 2 \alpha_0 ) \biggr\}                             \nonumber  \\
 & = &  { 1 \over 4 \sin ( 2 \alpha_0 ) }
           \biggl[  1 - \frac{  \cos ( 2 \alpha_0 ) \left\{  X_1^2 + X_2^2 + 2 \cos (  2 \alpha_0 ) \right\} } 
                 { 2  \sin^2 ( 2 \alpha_0 ) } \left\{   \cos ( 4 \beta + 2 \phi ) + 1 \right\}    \biggr.   \nonumber  \\
  & &  \hskip2.2cm  + \biggl.  \frac{ X_2  \sqrt{ X_1^2 + X_2^2 + 2 \cos (  2 \alpha_0 ) } }
                                   { \sin ( 2 \alpha_0 ) }
                                \left\{   \sin ( 4 \beta +  \phi -2 \alpha_0) -  \sin ( \phi + 2 \alpha_0) \right\}  \biggr] 
                                  \nonumber  \\
  & = &  { 1 \over 4 \sin ( 2 \alpha_0 ) }
           \biggl[  1          - \frac{  \cos ( 2 \alpha_0 ) }{ 2  \sin^2 ( 2 \alpha_0 ) } 
          \frac{ \vert m_{ee} \vert^2 }{ \vert m_{ee}^{(1)} \vert  \vert m_{ee}^{(2)} \vert }
           \left\{   \cos ( 4 \beta + 2 \phi ) + 1 \right\}                    \biggr.                     \nonumber  \\
   & &  \hskip2.2cm  + \biggl.  \frac{ 1 }{ \sin ( 2 \alpha_0 ) }
                              \frac{ \vert m_{ee} \vert }{ \vert m_{ee}^{(1)} \vert }
                                \left\{   \sin ( 4 \beta +  \phi -2 \alpha_0) -  \sin ( \phi + 2 \alpha_0) \right\}  \biggr].
  \label{ccbd}
\end{eqnarray}
The coefficient $\alpha_1$, Eq.(\ref{ccan}), involves the cosine function of $\beta$
   with the period $ \pi $, and the coefficient $\alpha_2$ involves the cosine (sine)
    function of $\beta$ with the period $ \pi /2$.
The $\alpha$ is the sum of the constant $\alpha_0$, the term involving the cosine function
   of $\beta$ with the period $ \pi $, and the other term involving the cosine (sine) function
   of $\beta$ with the period $ \pi/2$.
If needed, the higher orders of $ O(X_3^n) $ can be calculated in the same manner.
\subsubsection{  The third case;
        $  \vert m_{ee}^{(2)} \vert >   \vert m_{ee}^{(3)} \vert  >>  \vert m_{ee}^{(1)} \vert   $.}
The third case is that the value of the lightest neutrino mass $m_1$
    (= $ \vert m_{ee}^{(1)} \vert /  \vert U_{e 1} \vert^2 $) is small enough to satisfy the
    following relation,
\begin{equation}
      \vert m_{ee}^{(2)} \vert >   \vert m_{ee}^{(3)} \vert >>  \vert m_{ee}^{(1)} \vert.
  \label{ccca}
\end{equation}
In this third case, $ \vert m_{ee}^{(1)} \vert $ is much smaller than 
   $  \vert m_{ee}^{(2)} \vert  $ and $  \vert m_{ee}^{(3)} \vert  $.
At first, let $ \vert m_{ee}^{(1)} \vert $ equal zero, and the difference of two Majorana
   phases $ \alpha - \beta $ satisfies,
\begin{equation}
   \vert m_{ee} \vert
    =   \vert \,  \vert m_{ee}^{(2)} \vert e^{2 i \alpha}
                +  \vert m_{ee}^{(3)} \vert e^{2 i \beta} \, \vert
    =    \vert \,  \vert m_{ee}^{(2)} \vert e^{2 i ( \alpha - \beta )}
                +  \vert m_{ee}^{(3)} \vert  \, \vert,
  \label{cccb}
\end{equation}
or \cite{rf:Rod3} 
\begin{equation}
 \cos ( 2 ( \alpha - \beta ) ) = \frac{  \vert m_{ee} \vert^2 -  \vert m_{ee}^{(2)} \vert^2 - \vert m_{ee}^{(3)} \vert^2 }
                                  { 2 \vert m_{ee}^{(2)} \vert  \vert m_{ee}^{(3)} \vert }.
  \label{cccc}
\end{equation}
In Ref.\cite{rf:Rod3}, the dependence of the effective Majorana mass $\vert m_{ee} \vert$
   on $ \alpha - \beta $, Eq.(\ref{cccc}), was studied in detail.
Defining this difference $ \alpha - \beta $ as $\gamma_0$, we have
\begin{equation}
   \alpha =  \gamma_0 + \beta ,
  \label{cccd}
\end{equation}
and
\begin{equation}
    \cos ( 2 \gamma_0 ) = \frac{  \vert m_{ee} \vert^2 -  \vert m_{ee}^{(2)} \vert^2 - \vert m_{ee}^{(3)} \vert^2 }
                                  { 2 \vert m_{ee}^{(2)} \vert  \vert m_{ee}^{(3)} \vert }.
  \label{ccce}
\end{equation}
In practice, the $ \vert m_{ee}^{(1)} \vert $ is not zero, and we will obtain the isocontour
   of $ \vert m_{ee} \vert $ by the method of perturbation with $ \vert m_{ee}^{(1)} \vert $,
   or more formally with 
   $  \vert m_{ee}^{(1)} \vert  /   \sqrt{  \vert m_{ee}^{(3)} \vert  \vert m_{ee}^{(2)} \vert }
          \equiv Y_1  $,
  which is discussed in Appendix.
When one takes up to the first order of $Y_1$, the relation between $\alpha$ and $\beta$
   becomes (see Eq.(\ref{op}) in Appendix),
\begin{eqnarray}
  \alpha &=& \gamma_0 + \beta +  Y_1 { 1 \over 2 \sin ( 2 \gamma_0 ) }
                    \sqrt { Y_3^2 + Y_2^2 + 2 \cos (  2 \gamma_0 ) } \,  
                    \cos ( 2 \beta - \varphi ) + O(Y_1^2)                  \nonumber  \\
            &=&  \gamma_0 + \beta +  \left( {m_1 \over m_3 m_2} \right)
                    \frac{  \vert m_{ee} \vert }{ 2 \sin ( 2 \gamma_0 ) }    
                    \frac{ \vert  U_{e 1} \vert^2 }{ \vert U_{e 3} \vert^2   \vert U_{e 2} \vert^2 }   
                     \cos ( 2 \beta - \varphi )  + O(Y_1^2),
  \label{cccf}
\end{eqnarray}
where
\begin{equation}
    Y_2 \equiv \sqrt{ \vert m_{ee}^{(2)} \vert \over  \vert m_{ee}^{(3)} \vert }, \hskip1cm
   Y_3 \equiv \sqrt{ \vert m_{ee}^{(3)} \vert \over  \vert m_{ee}^{(2)} \vert }, \hskip1cm
   \tan \varphi = - \frac{ Y_2 \sin ( 2 \gamma_0 ) }{Y_3 +  Y_2 \cos ( 2 \gamma_0 ) }.
  \label{cccg}
\end{equation}
The $\alpha$ is the sum of the constant $\gamma_0$, the $\beta$, and the cosine
   function of $\beta$ with the period $\pi$ which function is multiplied by the
   small quantity.
As discussed in Appendix, the approximation using 
  the expansion in $Y_1$ does not work near the
   maximum value $  \vert m_{ee} \vert_{\rm max} $ or the minimum value
   $   \vert m_{ee} \vert_{\rm min} $.
   
In subsection 3.3, we have considered the three cases and developed the
   appropriate method of perturbation in each case.
However, the case, 
   $ \vert m_{ee}^{(2)} \vert  >  \vert m_{ee}^{(1)} \vert
          \mathrel{ \rlap{\raise 0.511ex \hbox{$>$}}{\lower 0.511ex \hbox{$\sim$}}}
          \vert m_{ee}^{(3)} \vert $,
   has not been discussed.
Unfortunately, we are not able to find a useful method of approximation in this case.
%
%
%
%
%
%
%
%
%
%
\section{Majorana phases for the inverted mass ordering}
In this chapter, we derive the equation representing the isocontour of the effective
   Majorana mass $ \vert m_{ee} \vert $ in the $ \beta \alpha-$plane in the inverted
   mass ordering case ($ m_2 >m_1>m_3 $).
As discussed in section 2, the lightest neutrino mass $m_3$ is treated as a free parameter,
   and when we set the value of $m_3$, the effective Majorana mass $ \vert m_{ee} \vert $
   is considered as a function of $ \beta $ and $ \alpha $.
If the value of $ \vert m_{ee} \vert $ is determined, the constraint on $ \beta $ and
   $ \alpha $ is obtained, yet we can not obtain both values of $ \beta $ and $ \alpha $
   simultaneously as in the case of the normal mass ordering.
In the inverted mass ordering case, the order of three
   $ \vert m_{ee}^{(i)} \vert $ is distinctly known for an arbitrary value of $m_3$,
\begin{equation}
   \vert m_{ee}^{(1)} \vert >   \vert m_{ee}^{(2)} \vert  >>  \vert m_{ee}^{(3)} \vert,
   \hskip1cm  m_1 \approx m_2.
  \label{dx}
\end{equation}
Hence, we can discuss about the isocontour of $ \vert m_{ee} \vert $ in short
   compared to the normal mass ordering case.
In the same manner as the normal mass ordering case, we first obtain the isocontour of
   $ \vert m_{ee} \vert $ around the point of maximum $ \vert m_{ee} \vert $
   in the $ \beta \alpha-$plane, and next those around the point of minimum
   $ \vert m_{ee} \vert $, restricting the region of $ \beta $ and $ \alpha $
   as Eq.(\ref{cx}).
In the final subsection, we obtain the isocontour except around the point of maximum
   or minimum $ \vert m_{ee} \vert $.
\subsection{Around maximum value of $ \vert m_{ee} \vert $.}
For an arbitrary value of the lightest neutrino mass $m_3$, the effective Majorana mass
   $ \vert m_{ee} \vert $ takes a maximum value,
\begin{equation}
    \vert m_{ee} \vert_{\rm max}
     =  \vert m_{ee}^{(1)} \vert +  \vert m_{ee}^{(2)} \vert
          +  \vert m_{ee}^{(3)} \vert,
  \label{daa}
\end{equation}
at the point $ (\beta, \alpha) = (0,0) $ in the $ \beta \alpha-$plane.
Let us expand $ \vert m_{ee} \vert^2 $ around this point.
Including the terms up to second order on the two variables
   $ \beta $ and $ \alpha $ in the expansion, we have the same expressions as those from
   Eq.(\ref{cac}) to Eq.(\ref{cak}) in section 3.
If one neglects the fourth order of two variables $ \beta $ and $ \alpha $ or higher orders,
   the isocontour around the point of maximum $ \vert m_{ee} \vert_{\rm max} $ becomes
   the ellipse represented by   
\begin{equation}
   \frac{ \beta''^2 }{ \left( \sqrt{ J \over \lambda_-  } \right)^2 }
   +  \frac{ \alpha''^2 }{ \left( \sqrt{ J \over \lambda_+  } \right)^2 } = 1,
   \label{dab}
\end{equation}
where
\begin{equation}
    \left(     \begin{array}{c}
         \beta''            \\
         \alpha''
                 \end{array}
                                 \right)
     =
     \left(     \begin{array}{cc}
        \cos \theta   &   \sin \theta         \\
        - \sin \theta  &  \cos \theta
                 \end{array}
                                 \right)
     \left(     \begin{array}{c}
         \beta            \\
         \alpha
                 \end{array}
                                 \right),
  \label{dac}
\end{equation}
and
\begin{equation}
     J  =  {1 \over 4}  \biggl\{   \vert m_{ee} \vert_{\rm max}^2 -  \vert m_{ee} \vert^2
                                                                                                                   \biggr\} > 0.
  \label{dad}
\end{equation}
The direction of the $\beta''$ axis is produced by a counterclockwise rotation of the
   $\beta$ axis by the angle $\theta (>0)$.

Since $ \vert m_{ee}^{(3)} \vert $ is much smaller than $ \vert m_{ee}^{(1)} \vert $ and
   $ \vert m_{ee}^{(2)} \vert $, one can neglect the terms of order
   $ O(\vert m_{ee}^{(3)} \vert^2) $ in the expressions of Eq.(\ref{cag}) in the inverted
   mass ordering case.
The $\tan \theta$ and the ratio of the minor axis to the major axis,
   $ \sqrt{ \lambda_{-} / \lambda_{+} } $, are approximated, respectively,
\begin{eqnarray}
   \tan \theta  & \approx & \frac{  \vert m_{ee}^{(3)} \vert }{  \vert m_{ee}^{(1)} \vert } \,
                                            = \, { m_3 \over m_1 } \frac{ \vert U_{e 3} \vert^2 }{ \vert U_{e 1} \vert^2 } << 1,
                                                                       \nonumber  \\
    \sqrt{ \lambda_{-} \over \lambda_{+} }
     & \approx &  \sqrt{ \frac{ (  \vert m_{ee}^{(1)} \vert +  \vert m_{ee}^{(2)} \vert ) }
                                {   \vert m_{ee}^{(1)} \vert   \vert m_{ee}^{(2)} \vert }   \vert m_{ee}^{(3)} \vert  } 
    \approx \sqrt{ m_3 \over m_1 }
       \sqrt{ \frac{ \vert U_{e 3} \vert^2 ( 1 -   \vert U_{e 3} \vert^2 ) }{ \vert U_{e 1} \vert^2  \vert U_{e 2} \vert^2 } },
   \label{dae}
\end{eqnarray}
where we have used $ m_1 \approx m_2 $.
We should notice here that, in the inverted (normal) mass ordering case,
  the lightest neutrino mass $m_3 \, (m_1) $ is regarded as a free parameter.
When the value of the lightest neutrino mass $m_3$ is decreased, both the angle
   $\theta$ and the ratio $ \sqrt{ \lambda_{-} / \lambda_{+} } $ decrease monotonously.
In the $ m_3 \rightarrow 0 $ limit, we have
\begin{equation}
   \theta  \rightarrow   0,   \hskip1cm
   \sqrt{ \lambda_{-} \over \lambda_{+} }   \rightarrow   0,  \hskip1cm  (m_3 \rightarrow 0).
  \label{dag}
\end{equation}
The behavior of the angle $\theta$ and the ratio $ \sqrt{  \lambda_{-} /  \lambda_{+} } $
   by decreasing the value of the lightest neutrino mass in this subsection of the inverted
   mass ordering case is strikingly different from that in the subsection 3.1 of the normal
   mass ordering case.
When the size of the major axis of the ellipse is a quarter of $ \pi$, 
   $ 2 \sqrt{  J /  \lambda_{-} } =  \pi/4 $, the effective Majorana mass becomes
\begin{equation}
      \vert m_{ee} \vert   \approx    \vert m_{ee} \vert_{\rm max} 
                                           - {\pi^2 \over 32}   \vert m_{ee}^{(3)} \vert.
  \label{dah}
\end{equation}
\subsection{Around minimum value of $  \vert m_{ee} \vert  $.}
Owing to the relation Eq.(\ref{dx}) in the inverted mass ordering case, the effective
   Majorana mass takes a minimum value,
\begin{equation}
    \vert m_{ee} \vert_{\rm min}
     =  \vert m_{ee}^{(1)} \vert -  \vert m_{ee}^{(2)} \vert
          -  \vert m_{ee}^{(3)} \vert,
  \label{dba}
\end{equation}
at the point $ (\beta, \alpha) = ({ \pi / 2 },  - { \pi / 2 }) $.
Let us expand $ \vert m_{ee} \vert^2 $ around this point.
Including the terms up to second order in the expansion, we have the same expressions
   as those from Eq.(\ref{cbab}) to Eq.(\ref{cbai}) in section 3.
If one disregards the fourth order or higher orders, the isocontour around the point of
   minimum $ \vert m_{ee} \vert_{\rm min} $ becomes the ellipse represented by
\begin{equation}
   \frac{  \beta''^2 }{ \left( \sqrt{ K \over \lambda_-  } \right)^2 }
   +  \frac{  \alpha''^2 }{ \left( \sqrt{ K \over \lambda_+  } \right)^2 } = 1,
   \label{dbb}
\end{equation}
where
\begin{equation}
    \left(   \begin{array}{c}
                \beta''          \\
                \alpha''
             \end{array}
                                 \right)
  =  \left(     \begin{array}{cc}
        \cos \theta   &   \sin \theta         \\
        - \sin \theta  &  \cos \theta
                 \end{array}
                                 \right)
     \left(   \begin{array}{c}
                   \beta'           \\
                   \alpha'
               \end{array}
                                 \right)
  =
     \left(     \begin{array}{cc}
        \cos \theta   &   \sin \theta         \\
        - \sin \theta  &  \cos \theta
                 \end{array}
                                 \right)
     \left(     \begin{array}{c}
         ( \beta - { \pi \over 2 })            \\
         ( \alpha + { \pi \over 2 })
                 \end{array}
                                 \right),
  \label{dbc}
\end{equation}
and
\begin{equation}
     K   =  {1 \over 4}  \biggl\{   \vert m_{ee} \vert^2 - \vert m_{ee} \vert_{\rm min}^2
                                                                                                        \biggr\} > 0.
  \label{dbd}
\end{equation}
This ellipse has the center at $ (\beta, \alpha) =  ({ \pi / 2 },  - { \pi / 2 }) $, and
   the direction of the $\beta''$ axis is produced by a clockwise rotation of the $\beta$
   axis by the angle $ \vert \theta \vert (>0)$.
   
Because one can neglect the terms of order $ O(\vert m_{ee}^{(3)} \vert^2) $ in the
   expressions of Eq.(\ref{cbah}) in the inverted mass ordering case, the angle $\theta$
   and the ratio of the minor axis to the major axis, $ \sqrt{ \lambda_{-} / \lambda_{+} } $,
   are approximated, respectively,
\begin{eqnarray}
   \theta  & \approx & - \frac{  \vert m_{ee}^{(3)} \vert }{  \vert m_{ee}^{(1)} \vert } \,
                                            = \, -{ m_3 \over m_1 } \frac{ \vert U_{e 3} \vert^2 }{ \vert U_{e 1} \vert^2 } < 0,  \hskip0.5cm   \vert \theta \vert << 1,
                                                                       \nonumber  \\
   \sqrt{ \lambda_{-} \over \lambda_{+} }
     & \approx &  \sqrt{ \frac{ (  \vert m_{ee}^{(1)} \vert - \vert m_{ee}^{(2)} \vert ) }
                                 {   \vert m_{ee}^{(1)} \vert   \vert m_{ee}^{(2)} \vert }  \vert m_{ee}^{(3)} \vert  }
   \approx \sqrt{ m_3 \over m_1 }
       \sqrt{ \frac{ \vert U_{e 3} \vert^2 ( \vert U_{e 1} \vert^2 - \vert U_{e 2} \vert^2 ) }{ \vert U_{e 1} \vert^2  \vert U_{e 2} \vert^2 } },
   \label{dbe}
\end{eqnarray}
where we have used $m_1 \approx m_2$.
When the value of the lightest neutrino mass $m_3$ is decreased, both
   the absolute value of the angle $ \vert \theta \vert $ and the ratio
   $ \sqrt{ \lambda_{-} / \lambda_{+} } $ decrease monotonously.
In the $  m_3 \rightarrow 0 $ limit, one has
\begin{equation}
   \theta  \rightarrow   0,   \hskip1cm
   \sqrt{ \lambda_{-} \over \lambda_{+} }   \rightarrow   0,  \hskip1cm  (m_3 \rightarrow 0).
  \label{dbg}
\end{equation}
The behavior of the angle $\theta$ and the ration $ \sqrt{  \lambda_{-} /  \lambda_{+} } $
   by decreasing the value of the lightest neutrino mass in this subsection of the inverted
   mass ordering case is strikingly different from that in the sub-subsection 3.2.1 of the
   normal mass ordering case.
When the size of the major axis of the ellipse is $ \pi / 4$, 
   the effective Majorana mass becomes
\begin{equation}
      \vert m_{ee} \vert
           \approx   \vert m_{ee} \vert_{\rm min} 
                      + {\pi^2 \over 32}   \vert m_{ee}^{(3)} \vert
            =  \vert m_{ee}^{(1)} \vert - \vert m_{ee}^{(2)} \vert 
                   - \left( 1- {\pi^2 \over 32} \right)  \vert m_{ee}^{(3)} \vert.
  \label{dbh}
\end{equation}
\subsection{ Region except extremal values of $  \vert m_{ee} \vert  $.}
In this subsection, we will find the isocontour not around the point of the maximum or
   minimum $  \vert m_{ee} \vert  $.
Since the relation
   $ \vert m_{ee}^{(1)} \vert >   \vert m_{ee}^{(2)} \vert  >>  \vert m_{ee}^{(3)} \vert $
   holds for an arbitrary value of the lightest neutrino mass $m_3$ in the inverted mass
   ordering case, we can use the method of perturbation with
   $  \vert m_{ee}^{(3)} \vert  /   \sqrt{  \vert m_{ee}^{(1)} \vert  \vert m_{ee}^{(2)} \vert }
      = X_3 << 1  $, which is used in section 3.3.1.
The relation between $\alpha$ and $\beta$ is, from Eq.(\ref{ccaq}),
\begin{eqnarray}
  \alpha  &=& \alpha_0 +  X_3 \,
                      { 1 \over 2 \sin ( 2 \alpha_0 ) }
                      \frac{ \vert m_{ee} \vert }
                      { \sqrt{  \vert m_{ee}^{(1)} \vert  \vert m_{ee}^{(2)} \vert } }
                        \,  \cos ( 2 \beta + \phi )      
                                + O(X_3^2)                               \nonumber  \\
            &=& \alpha_0 +  \left( {m_3 \over m_1 m_2} \right)
                    \frac{  \vert m_{ee} \vert }{ 2 \sin ( 2 \alpha_0 ) }    
                    \frac{ \vert  U_{e 3} \vert^2 }{ \vert U_{e 1} \vert^2   \vert U_{e 2} \vert^2 }   
                     \cos ( 2 \beta + \phi )  + O(X_3^2),
    \label{dca}
\end{eqnarray}
where
\begin{equation}
  \tan \phi = - \frac{ X_2 \sin ( 2 \alpha_0 ) }{X_1 +  X_2 \cos ( 2 \alpha_0 ) },
  \hskip0.5cm  X_1 = \sqrt{ \vert m_{ee}^{(1)} \vert
                             \over  \vert m_{ee}^{(2)} \vert }, \hskip0.5cm
  X_2 = \sqrt{ \vert m_{ee}^{(2)} \vert \over  \vert m_{ee}^{(1)} \vert }.
  \label{dcc}
\end{equation}
The $ \alpha_0 $ is given by \cite{rf:Rod3}
\begin{equation}
 \cos ( 2 \alpha_0 ) = \frac{  \vert m_{ee} \vert^2 -  \vert m_{ee}^{(1)} \vert^2 - \vert m_{ee}^{(2)} \vert^2 }
                                  { 2 \vert m_{ee}^{(1)} \vert  \vert m_{ee}^{(2)} \vert }.
  \label{dcb}
\end{equation}
Note that this perturbative calculation can not be used near the point of the maximum
   $ \vert m_{ee} \vert_{\rm max} $ or the minimum $ \vert m_{ee} \vert_{\rm min} $
   as discussed in section 3.3.1.
   
When the mass $m_3$ decreases, the factor $  m_3 / (m_1 m_2) $ in the coefficient
   of $ \cos (2 \beta + \phi) $ in Eq.(\ref{dca}) decreases.
Hence, the contribution of the second term in the right-handed side of Eq.(\ref{dca})
   becomes small when the value of the lightest neutrino mass $m_3$ decreases.
This behavior of the second term in the right-handed side of Eq.(\ref{dca}) by decreasing
   the value of the lightest neutrino mass in this subsection of the inverted mass ordering
   case is strikingly different from that in sub-subsection 3.3.1 of the normal mass
   ordering case.
If the $ m_3 \rightarrow 0 $ limit is taken, one has
\begin{equation}
    \alpha \rightarrow \alpha_0,  ~~  ( m_3 \rightarrow 0 ),
  \label{dcd}
\end{equation}
and $\alpha$ approaches the constant not depending on $\beta$ \cite{rf:Rod3}.
%
%
%
%
%
%
%
\section{Conclusion.}
In the $ \beta \alpha-$plane where $\beta$ and $\alpha$ are the Majorana
   CP-violating phases, we obtained analytically the equation representing the isocontour
   of the effective Majorana mass $ \vert m_{ee} \vert $ by the method of perturbation.
The effective Majorana mass $ \vert m_{ee} \vert $ is written as
\begin{equation}
  \vert m_{ee} \vert
   =  \left\vert \, \vert m_{ee}^{(1)} \vert +  \vert m_{ee}^{(2)} \vert e^{2 i \alpha}
    +  \vert m_{ee}^{(3)} \vert e^{2 i \beta} \, \right\vert,
  \label{ea}
\end{equation}
where $  \vert m_{ee}^{(1)} \vert  = m_1 \vert U_{e 1} \vert^2 $, 
   $  \vert m_{ee}^{(2)} \vert  = m_2 \vert U_{e 2} \vert^2  $, and
   $  \vert m_{ee}^{(3)} \vert  =  m_3 \vert U_{e 3} \vert^2 $.
The equation representing the isocontour of $ \vert m_{ee} \vert $ is expressed
   by the following six quantities;
   $ \vert m_{ee} \vert $,
   two lepton mixing angles $(\theta_{12},  \theta_{13})$,
   two neutrino mass squared differences
   $ \bigtriangleup m_{\odot}^2 $ (responsible for the solar neutrino oscillation) and
   $ \bigtriangleup  m_{\rm A}^2 $ (responsible for the atmospheric neutrino oscillation),
   and the lightest neutrino mass which is regarded as a free parameter.
We studied how the isocontour of $ \vert m_{ee} \vert $ changes in the
   $ \beta \alpha-$plane when the value of the lightest neutrino mass is varied in the case
   of the normal mass ordering and the inverted mass ordering, respectively.

In the normal mass ordering case $(m_3 >m_2>m_1)$, we assumed the relation
   $ \vert m_{ee}^{(2)} \vert  >   \vert m_{ee}^{(3)} \vert $.
When the value of the lightest neutrino mass $m_1$ is varied, the
   $ \vert m_{ee}^{(1)} \vert $ changes from zero to the maximum of
   $ \vert m_{ee}^{(i)} \vert (i=1,2,3) $.
The effective Majorana mass $ \vert m_{ee} \vert $ has the maximum value at the point
    $ (\beta, \alpha) = (0,0) $, and the isocontour of $ \vert m_{ee} \vert $ around this point
    is the ellipse whose major axis is tilted in the counterclockwise direction of the $\beta$
    axis on the $ \beta \alpha-$plane.
When the value of the lightest neutrino mass $m_1$ is varied, we examined how the
   form of the ellipse or the direction of the major axis changes in the $ \beta \alpha-$plane
   as discussed in section 3.1.
In finding the minimum value of $ \vert m_{ee} \vert $, it is convenient to divide into
   the following three regions according to the size of $\vert m_{ee}^{(1)} \vert$,
\begin{itemize}
  \item  region (A);     $ \hskip1cm   \vert m_{ee}^{(2)} \vert +  \vert m_{ee}^{(3)} \vert <  \vert m_{ee}^{(1)} \vert $,
  \item  region (B);   $ \hskip1cm   \vert m_{ee}^{(2)} \vert -   \vert m_{ee}^{(3)} \vert
                                    <  \vert m_{ee}^{(1)} \vert <  \vert m_{ee}^{(2)} \vert +  \vert m_{ee}^{(3)} \vert $,
   \item  region (C);    $ \hskip1cm   \vert m_{ee}^{(1)} \vert <   \vert m_{ee}^{(2)} \vert -   \vert m_{ee}^{(3)} \vert $.
\end{itemize}
In the region (A), the $ \vert m_{ee} \vert $ has the minimum value at the point
   $ (\beta, \alpha) = ({ \pi / 2 },  - { \pi / 2 }) $.
The isocontour of $ \vert m_{ee} \vert $ around this point is the ellipse whose major axis is
   in the clockwise direction of the $\beta$ axis.
On the boundary between the region (A) and region (B),
   $  \vert m_{ee}^{(1)} \vert =  \vert m_{ee}^{(2)} \vert  +  \vert m_{ee}^{(3)} \vert $,
   the effective Majorana mass $ \vert m_{ee} \vert $ takes a minimum value at the point
   $ (\beta, \alpha) = ({ \pi / 2 },  - { \pi / 2}) $.
Around this point, the isocontour of $ \vert m_{ee} \vert $ becomes the straight line
   within the approximation up to the second order.
If one takes into account the higher order terms such as 
   $ O(\beta - \pi/2)^4) $, etc., the isocontour would become a closed line.
In the region (B), the $ \vert m_{ee} \vert $ has the minimum value zero at the point
   $ (\beta_c, \alpha_c) $ where $ 0   <  \beta_c  <  {\pi / 2} $
   and $ - {\pi / 2}   <  \alpha_c  <  - {\pi / 4} $.
The isocontour around this point is the ellipse.
On the boundary between the region (B) and region (C),
   $  \vert m_{ee}^{(1)} \vert =  \vert m_{ee}^{(2)} \vert  -  \vert m_{ee}^{(3)} \vert $,
   the $ \vert m_{ee} \vert $ has the minimum value at the point
   $ (\beta, \alpha) = (0,  - { \pi / 2}) $.
Around this point, the isocontour of $ \vert m_{ee} \vert $ becomes the straight line
   within the approximation up to the second order.
If we take into account the higher order terms such as 
   $ O(\beta^4) $, etc., the isocontour would become a closed line.
In the region (C), the $ \vert m_{ee} \vert $ takes the minimum value at the point
   $ (\beta, \alpha) = (0,  - { \pi / 2 }) $.
The isocontour of $ \vert m_{ee} \vert $ around this point is the ellipse whose major axis is
   in the counterclockwise direction of the $\beta$ axis.
We also found the isocontour not around the point of the maximum or minimum
   $ \vert m_{ee} \vert $ in the $ \beta \alpha-$plane.
The equation representing such isocontour is obtained analytically using the
   approximation method in the following three cases, (a)$\sim$(c). \\
$\cdot$ (a)  $ \vert m_{ee}^{(1)} \vert >   \vert m_{ee}^{(2)} \vert >>
    \vert m_{ee}^{(3)} \vert $.  \\
In this case, the $\alpha$ is the sum of the constant term not depending on $\beta$,
   and the term of the cosine function of $\beta$ with the period $\pi$ which function
   is multiplied by the small quantity.
When the value of the lightest neutrino mass $m_1$ is decreased, the contribution of the
   term involving the cosine function of $\beta$ becomes larger.
While the value of $\beta$ can not be determined  at all, the value of $\alpha$ can be
   restricted within a rather narrow range.\\
$\cdot$ (b) $\vert m_{ee}^{(1)} \vert > \vert m_{ee}^{(2)} \vert > 
          \vert m_{ee}^{(3)} \vert $. \\
The $\alpha$ is the sum of the constant term not depending on $\beta$, the term
   proportional to the cosine function of $\beta$ with the period $\pi$, and the term
   proportional to the cosine (sine) function of $\beta$ with the period $\pi / 2$.
While the value of $\beta$ can not be determined  at all, the value of $\alpha$ can be
   restricted within a certain range.
When the value of the lightest neutrino mass $m_1$ is decreased, the range restricting
   the value of $\alpha$ widens.\\
$\cdot$ (c)  $  \vert m_{ee}^{(2)} \vert >   \vert m_{ee}^{(3)} \vert  >>
                        \vert m_{ee}^{(1)} \vert $. \\
The $\alpha$ is the sum of the constant term not depending on $\beta$, the $\beta$,
   and the term proportional to the cosine function of $\beta$ with the period $\pi$.
When the value of the lightest neutrino mass $m_1$ is decreased, 
   the term involving the cosine function of $\beta$ approaches zero.
Both the value of $\beta$ and that of $\alpha$ can be restricted within certain ranges,
   respectively.
   
A thing which has not been discussed is the case,
    $ \vert m_{ee}^{(2)} \vert  >   \vert m_{ee}^{(1)} \vert
          \mathrel{ \rlap{\raise 0.511ex \hbox{$>$}}{\lower 0.511ex \hbox{$\sim$}}}
          \vert m_{ee}^{(3)} \vert $,
    where the orders of $ \vert m_{ee}^{(1)} \vert $, $  \vert m_{ee}^{(2)} \vert $, and
    $ \vert m_{ee}^{(3)} \vert  $ are equal in magnitude.
Unfortunately, we can not find a useful method of approximation in this case.

In the inverted mass ordering case $(m_2 >m_1>m_3)$, the following relation holds for
   an arbitrary value of the lightest neutrino mass $m_3$,
\begin{equation}
   \vert m_{ee}^{(1)} \vert >   \vert m_{ee}^{(2)} \vert  >>  \vert m_{ee}^{(3)} \vert,
   \hskip1cm   m_1 \approx m_2.
  \label{ef}
\end{equation}
The effective Majorana mass $ \vert m_{ee} \vert $ takes the maximum value at the point
   $ (\beta, \alpha) = (0,0) $, and the isocontour of $ \vert m_{ee} \vert $ around this
   point is the ellipse whose major axis is slightly tilted in the counterclockwise direction
   of the $\beta$ axis on the $ \beta \alpha-$plane.
When the value of the lightest neutrino mass $m_3$ is decreased, the major axis of the
   ellipse approaches the $\beta$ axis, and the ratio of the minor axis of the ellipse to
   the major axis approaches zero.
The $ \vert m_{ee} \vert $ has the minimum value at the point
    $ (\beta, \alpha) = (\pi/2, -\pi/2) $, and the isocontour of $ \vert m_{ee} \vert $ around
    this point is the ellipse whose major axis is slightly
    tilted in the clockwise direction of the $\beta$
    axis on the $ \beta \alpha-$plane.
When the value of the lightest neutrino mass $m_3$ is decreased, the major axis of the
   ellipse approaches the $\beta$ axis, and the ratio of the minor axis of the ellipse to
   the major axis approaches zero.
We also found the isocontour not around the point of the maximum or minimum
   $ \vert m_{ee} \vert $ in the $ \beta \alpha-$plane.
The equation representing that isocontour is such that 
   the $\alpha$ is the sum of the constant term not depending on $\beta$,
   and the term of the cosine function of $\beta$ with the period $\pi$ which function
   is multiplied by the small quantity.
When the value of the lightest neutrino mass $m_3$ is decreased, the contribution of the
   term involving the cosine function of $\beta$ becomes small.
While the value of $\beta$ can not be determined  at all, the value of $\alpha$ can be
   restricted within a very narrow range.
   
So far the constraints between the Majorana CP-violating phases $\alpha$ and $\beta$
   have been investigated mainly by the numerical calculations.
At present, we can not decide whether the mass of the neutrino obeys the normal mass
   ordering or the inverted mass ordering, and the absolute neutrino mass scale is
   unknown.
Under such circumstances remaining numerous possibilities, we would like to see
   prospectively the constraints between $\alpha$ and $\beta$ (the isocontour of
   $ \vert m_{ee} \vert $ in the $ \beta \alpha-$plane).
In this paper, we obtained the equation of the constraint between $\alpha$ and
   $\beta$ analytically
   by the method of perturbation, and clarified how the constraint changes in
   the $ \beta \alpha-$plane when the value of the lightest neutrino mass is varied.
Since the perturbative calculations are possible up to any order, one can analytically
   calculate according to the required accuracy in principle.

A positive signal of the neutrinoless double beta decay has not been detected in the
   experiments, and the upper limit of the effective Majorana mass $ \vert m_{ee} \vert $
   has been reported.
The KamLAND-Zen experiment provides the upper limit \cite{rf:Gan},
\begin{equation}
       \vert m_{ee} \vert  < (120-250) \, {\rm meV} \hskip0.5cm  {\rm at ~~ 90 \% ~ C.L.}
  \label{eg}
\end{equation}
We expect that the upper limit of $ \vert m_{ee} \vert $ will be lowered by the future
   neutrinoless double beta decay experiments.
When the upper limit of $ \vert m_{ee} \vert $ is put at a certain value, one can exclude
   some parameter regions in the $ \beta \alpha-$plane for a given value of the
   lightest neutrino mass.
As discussed in section 2, we regard the other four parameters,
   $\theta_{12}, \theta_{13}$, 
   $ \bigtriangleup m_{\odot}^2 $, and $ \bigtriangleup  m_{\rm A}^2 $ as given
   quantities in this paper.
Let us assume that the value of the upper limit of $ \vert m_{ee} \vert $ becomes so small
   that the inverted mass ordering is denied \cite{rf:PasPet}.
In the normal mass ordering, we consider here the case that the lightest neutrino mass
   $m_1$ satisfies the relation,
   $  \vert m_{ee}^{(1)} \vert >   \vert m_{ee}^{(2)} \vert >>  \vert m_{ee}^{(3)} \vert  $.
If the upper limit of $ \vert m_{ee} \vert $ becomes
\begin{equation}
      \vert m_{ee} \vert
          \le   \vert m_{ee} \vert_{\rm max} 
                  - {\pi^2 \over 32}   \vert m_{ee}^{(3)} \vert
          =  \vert m_{ee}^{(1)} \vert + \vert m_{ee}^{(2)} \vert 
                + \left( 1- {\pi^2 \over 32} \right)  \vert m_{ee}^{(3)} \vert,
  \label{eh}
\end{equation}
then the parameter region inside the ellipse, Eq.(\ref{caq}),
\begin{equation}
   \frac{\beta''^2}{ \left(  \pi \over 8 \right)^2 }
   + \frac{\alpha''^2}
        { \left( { \pi \over 8 } \sqrt{  \frac{ (  \vert m_{ee}^{(1)} \vert +  \vert m_{ee}^{(2)} \vert ) }
                          {   \vert m_{ee}^{(1)} \vert   \vert m_{ee}^{(2)} \vert }   \vert m_{ee}^{(3)} \vert } 
                \,  \right)^2 }
   = 1,
     \label{ei}
\end{equation}
is excluded in the $ \beta \alpha-$plane.
Furthermore, if the upper limit of $ \vert m_{ee} \vert $ is lowered as
\begin{equation}
     \vert m_{ee} \vert  \le  \sqrt{  \vert m_{ee}^{(1)} \vert^2 +  \vert m_{ee}^{(2)} \vert^2 },
  \label{ej}
\end{equation}
then, from Eq.(\ref{ccab}) and Eq.(\ref{ccar}), the following parameter region,
\begin{eqnarray}
  & &  - { \pi \over 4 } 
                    - \frac{ \sqrt{  \vert m_{ee}^{(1)} \vert^2 +  \vert m_{ee}^{(2)} \vert^2 }  }
                      { 2 \vert m_{ee}^{(1)} \vert  \vert m_{ee}^{(2)} \vert }
                       \vert m_{ee}^{(3)} \vert  \,  \cos ( 2 \beta + \phi )      
           \le  \alpha                                                           \nonumber \\
  & &  \le   { \pi \over 4 } 
                    + \frac{ \sqrt{  \vert m_{ee}^{(1)} \vert^2 +  \vert m_{ee}^{(2)} \vert^2 }  }
                      { 2 \vert m_{ee}^{(1)} \vert  \vert m_{ee}^{(2)} \vert }
                       \vert m_{ee}^{(3)} \vert  \,  \cos ( - 2 \beta + \phi ),      
   \label{ek}
\end{eqnarray}
is excluded.
In the future, if the neutrinoless double beta decay is detected by experiments and we
   can confirm the neutrino is Majorana fermion, a next challenge is to restrict the values
   of the Majorana phases $\alpha$ and $\beta$.
\newpage
%
%
%
\noindent{\Large\bf Appendix}
\appendix 
%
%
\section*{Expansion in $Y_1$}
\renewcommand{\theequation}{A.\arabic{equation}}
\setcounter{equation}{0}
The effective Majorana mass $  \vert m_{ee} \vert  $ is written as
\begin{equation}
   \vert m_{ee} \vert
 =  \vert \, \vert m_{ee}^{(1)} \vert +  \vert m_{ee}^{(2)} \vert e^{2 i \alpha}
   +  \vert m_{ee}^{(3)} \vert e^{2 i \beta} \, \vert.
  \label{oa}
\end{equation}
When the value of $  \vert m_{ee} \vert  $ is given, we will obtain the constraint equation
   between $\alpha$ and $\beta$ under the following condition,
\begin{equation}
   \vert m_{ee}^{(2)} \vert >  \vert m_{ee}^{(3)} \vert  >>  \vert m_{ee}^{(1)} \vert.
  \label{ob}
\end{equation}
The effective mass $  \vert m_{ee} \vert  $ can be rewritten as
\begin{eqnarray}
   \vert m_{ee} \vert
 &=&  \vert \, \vert m_{ee}^{(3)} \vert +  \vert m_{ee}^{(2)} \vert e^{2 i \gamma}
   +  \vert m_{ee}^{(1)} \vert e^{-2 i \beta} \, \vert   \nonumber \\
 & \equiv & \sqrt{ \vert m_{ee}^{(3)} \vert  \vert m_{ee}^{(2)} \vert } ~
       \vert Y_3 + Y_2 \, e^{2 i \gamma}+Y_1 \, e^{ - 2 i \beta} \vert,
  \label{oc}
\end{eqnarray}
where $ \gamma \equiv \alpha - \beta $ and
\begin{equation}
   Y_3 \equiv \sqrt{ \vert m_{ee}^{(3)} \vert \over  \vert m_{ee}^{(2)} \vert }, \hskip1cm
   Y_2 \equiv \sqrt{ \vert m_{ee}^{(2)} \vert \over  \vert m_{ee}^{(3)} \vert },\hskip1cm
   Y_1 \equiv   \frac{ \vert m_{ee}^{(1)} \vert  } { \sqrt{ \vert m_{ee}^{(3)} \vert  \vert m_{ee}^{(2)} \vert } }.
     \label{od}
\end{equation}
They satisfy, $ 1 >> Y_1$, $ Y_2 > Y_3 >> Y_1 $,   $~~ Y_3  Y_2 =1$, and
\begin{equation}
   \frac{ \vert m_{ee} \vert^2 }{ \vert m_{ee}^{(3)} \vert  \vert m_{ee}^{(2)} \vert }
     =  \vert Y_3 + Y_2 \, e^{2 i \gamma} + Y_1 \, e^{ - 2 i \beta} \vert^2.
  \label{oe}
\end{equation}

If we neglect $ Y_1$, we write $ \gamma = \gamma_0 $ and
\begin{equation}
   \frac{ \vert m_{ee} \vert^2 }{ \vert m_{ee}^{(3)} \vert  \vert m_{ee}^{(2)} \vert }
     =  \vert Y_3 + Y_2 \, e^{2 i \gamma_0}  \vert^2,
   \label{of}
\end{equation}
or
\begin{equation}
 \cos ( 2 \, \gamma_0 )
    = {1 \over 2} \left[  \frac{ \vert m_{ee} \vert^2 }{ \vert m_{ee}^{(3)} \vert  \vert m_{ee}^{(2)} \vert }
              - Y_3^2 - Y_2^2  \right].
  \label{og}
\end{equation}
Now, we will seek the $ \gamma $ by the perturbation with $ Y_1 $, and
   expand $ \gamma $ as
\begin{equation}
 \gamma  =   \sum_{n=0}^{\infty} \gamma_n Y_1^n 
                =  \gamma_0 + \gamma_1 Y_1 + \gamma_2 Y_1^2 + \cdots.
  \label{oh}
\end{equation}
Then, it leads
\begin{eqnarray}
   \frac{ \vert m_{ee} \vert^2 }{ \vert m_{ee}^{(3)} \vert  \vert m_{ee}^{(2)} \vert }
     & = & \vert  Y_3 + Y_2 \, e^{2 i ( \gamma_0 + \gamma_1 Y_1 + \gamma_2 Y_1^2+ \cdots ) }
                + Y_1 \, e^{ - 2 i \beta} \vert^2  \nonumber  \\
     & = &  \vert  b_0 + b_1 \, Y_1 + b_2 \, Y_1^2 + \cdots  \vert^2,
  \label{oi}
\end{eqnarray}
where
\begin{eqnarray}
  b_0  &\equiv&  Y_3+Y_2 \, e^{2 i \gamma_0},  \nonumber  \\
  b_1  &\equiv&  2 i \gamma_1 Y_2  \, e^{2 i \gamma_0} +  \, e^{ - 2 i \beta},
     \label{oj}
\end{eqnarray}
and so on.
The $b_0$ satisfies
\begin{equation}
  \frac{ \vert m_{ee} \vert^2 }{ \vert m_{ee}^{(3)} \vert  \vert m_{ee}^{(2)} \vert }
   = \vert b_0 \vert^2.
  \label{ok}
\end{equation}
When the order of $ Y_1^2 $ is neglected, one has
\begin{equation}
   \frac{ \vert m_{ee} \vert^2 }{ \vert m_{ee}^{(3)} \vert  \vert m_{ee}^{(2)} \vert }
     =  \vert  b_0 + b_1 \, Y_1 + O(Y_1^2)  \vert^2
     =  \vert b_0 \vert^2 + 2 {\rm Re} ( b_0^{\ast} b_1 ) Y_1 + O(Y_1^2).
  \label{ol}
\end{equation}
The $ \gamma_1$ can be determined by
\begin{equation}
  {\rm Re} ( b_0^{\ast} b_1 ) = 0,
  \label{om}
\end{equation}
and the result is
\begin{eqnarray}
  \gamma_1
     & = &   { 1 \over 2 \sin ( 2 \gamma_0 ) }
             \left\{ ( {\rm Re} \, b_0 ) \cos ( -2 \beta ) +  ( {\rm Im} \, b_0 ) \sin ( -2 \beta ) \right\}
                                                                                \nonumber \\
     & = &   { 1 \over 2 \sin ( 2 \gamma_0 ) } 
                     \left\{ \vert b_0 \vert  \,  \cos ( -2 \beta + \varphi ) \right\}       \nonumber \\
    & = &    { 1 \over 2 \sin ( 2 \gamma_0 ) }
                    \sqrt { Y_3^2 + Y_2^2 + 2 \cos (  2 \gamma_0 ) } \,  \cos ( -2 \beta + \varphi )
                                                                                                   \nonumber \\
   & = &  { 1 \over 2 \sin ( 2 \gamma_0 ) }
                 \frac{ \vert m_{ee} \vert }
                     { \sqrt{  \vert m_{ee}^{(3)} \vert  \vert m_{ee}^{(2)} \vert } }
                     \,  \cos ( -2 \beta + \varphi ),
  \label{on}
\end{eqnarray}
where
\begin{equation}
  \tan \varphi = - \frac{  {\rm Im} \, b_0 }{  {\rm Re} \, b_0 }
      = - \frac{ Y_2 \sin ( 2 \gamma_0 ) }{Y_3 +  Y_2 \cos ( 2 \gamma_0 ) }.
  \label{oo}
\end{equation}
Since $  \alpha - \beta = \gamma = \gamma_0 + \gamma_1 \, Y_1 + O(Y_1^2) $,
   we have
\begin{eqnarray}
  \alpha &=& \gamma_0 + \beta  +  Y_1 { 1 \over 2 \sin ( 2 \gamma_0 ) }
                    \sqrt { Y_3^2 + Y_2^2 + 2 \cos (  2 \gamma_0 ) } \, 
                     \cos ( 2 \beta - \varphi ) + O(Y_1^2)               \nonumber  \\
            &=&  \gamma_0 + \beta +  \left( {m_1 \over m_3 m_2} \right)
                    \frac{  \vert m_{ee} \vert }{ 2 \sin ( 2 \gamma_0 ) }    
                    \frac{ \vert  U_{e 1} \vert^2 }{ \vert U_{e 3} \vert^2   \vert U_{e 2} \vert^2 }   
                     \cos ( 2 \beta - \varphi )  + O(Y_1^2).
  \label{op}
\end{eqnarray}

In this expansion, it should be $ \sin 2 \gamma_0 \ne 0$.
This shows that we can not use this expansion near
   $ \gamma_0 = 0$ or $ \gamma_0 =  - \pi /2 $.
When $ \gamma_0 \approx 0$, it becomes $ \cos ( 2 \gamma_0 ) \approx 1$, and
   $  \vert m_{ee} \vert \approx  \vert m_{ee}^{(2)} \vert +  \vert m_{ee}^{(3)} \vert 
        \approx  \vert m_{ee} \vert_{\rm max} $.
When $ \gamma_0 \approx - \pi /2$, it becomes $ \cos ( 2 \gamma_0 ) \approx -1$, and
   $  \vert m_{ee} \vert \approx  \sqrt{ \vert m_{ee}^{(2)} \vert^2 
        +  \vert m_{ee}^{(3)} \vert^2 - 2  \vert m_{ee}^{(2)} \vert   \vert m_{ee}^{(3)} \vert }
       =  \vert m_{ee}^{(2)} \vert - \vert m_{ee}^{(3)} \vert \approx  
        \vert m_{ee} \vert_{\rm min} $.
From these, we understand that the expansion in $Y_1$ can not be used near the
   maximum value $  \vert m_{ee} \vert_{\rm max} $ or the minimum value
   $   \vert m_{ee} \vert_{\rm min} $.
%
%
%
%
%
\newpage
\end{document}